\documentclass[twocolumn]{aastex631}
\usepackage{graphicx}
\usepackage{bm}
\usepackage{amssymb}
\usepackage{amsmath}
\usepackage{units}
\usepackage{array}

\usepackage[T1]{fontenc}
\usepackage{aecompl}
\usepackage{calligra}
\usepackage{soul}
\DeclareMathAlphabet{\mathcalligra}{T1}{calligra}{m}{n}
\DeclareFontShape{T1}{calligra}{m}{n}{<->s*[2.2]callig15}{}

\newcommand{\Msun}[0]{M_\odot}
\newcommand{\Mdot}[0]{\dot M}

\newcommand{\edd}[0]{{\rm Edd}}

\newcommand{\crit}[0]{{\rm crit}}
\newcommand{\adv}[0]{{\rm adv}}

\newcommand{\ie}[0]{{\rm ie}}
\newcommand{\eg}[0]{{\rm e.g.,\,}} 
\newcommand{\synch}[0]{{\rm synch}}
\newcommand{\comp}[0]{{\rm comp}}
\newcommand{\brems}[0]{{\rm brems}}

\usepackage{booktabs}

\usepackage{orcidlink}

\usepackage[dvipsnames]{xcolor}

\submitjournal{ApJ}
\received{August 19, 2025}
\revised{October 17, 2025}
\accepted{October 25, 2025}

\shorttitle{BADAF}
\shortauthors{Tiede \& D'Orazio}
\graphicspath{{./}{Figures/}}

\begin{document}

\title{Hot, cold, and multi-component accretion flows around supermassive black hole binaries}

%
\author[0000-0002-3820-2404]{Christopher Tiede\,\orcidlink{0000-0002-3820-2404}}
\affiliation{Niels Bohr International Academy, Niels Bohr Institute, Blegdamsvej 17, 2100 Copenhagen, Denmark}
\email{christopher.tiede@nbi.ku.dk}
\author[0000-0002-1271-6247]{Daniel J. D'Orazio\,\orcidlink{0000-0002-1271-6247}}
\affiliation{Space Telescope Science Institute, 3700 San Martin Drive, Baltimore, MD 21218, USA}
\email{dorazio@stsci.edu}
\affiliation{Department of Physics and Astronomy, Johns
Hopkins University, 3400 North Charles Street, Baltimore,
Maryland 21218, USA}
\affiliation{Niels Bohr International Academy, Niels Bohr Institute, Blegdamsvej 17, 2100 Copenhagen, Denmark}

\begin{abstract}
\noindent
We develop a model for supermassive black hole binaries (SMBHBs) accreting below their Eddington limit, focusing on systems where hot, advection-dominated flows become viable. 
We specifically explore the spectral appearance of multi-component accretion flows where the solution can independently transition between cold, thin disks and hot, advection-dominated torii depending on the local accretion rate. 
Using a three-disk model, we compute spectral energy distributions for four possible accretion configurations and assess their observational signatures, including which frequencies might reflect variability at the binary orbital period. 
The spectral modeling reveals that binary accretion can self-consistently account for many of the properties of standard AGN, while the variability analysis shows that hydrodynamic modulation at the binary period is most likely in the thermal emission and low-frequency synchrotron components. 
Doppler boosting of emitting material bound to a single binary component would also induce periodic variability.
We apply our model to the SMBHB candidate PG1302-102 and demonstrate that a mixed-component accretion state (plus a jet feature) can self-consistently capture the observed broadband spectrum. 
Our model offers a framework for interpreting candidate SMBHBs and motivates future multi-wavelength follow-up of potential multi-messenger sources, as well as more detailed future modeling of multi-component binary accretion.
%
%
\end{abstract}


%
\section{Introduction} \label{S:intro}

Modern cosmology posits that nearly every massive galaxy hosts a supermassive black hole (SMBH) at its center, such that practically every major galaxy merger will create a supermassive black hole binary (SMBHB).
It is generally assumed that the majority of these SMBHBs will merge in the nascent galactic nucleus where they would produce the loudest gravitational waves (GWs) in the universe and contribute to the growth of SMBHs over cosmic time.
However, it remains unclear how exactly SMBHBs transition from galactic scales post-merger down to the sub-parsec separations required for gravitational radiation to merge the binary within a Hubble time \citep{Begel:Blan:Rees:1980}.
This uncertainty is exacerbated by the fact that at present there are no confirmed detections of a compact SMBHB that is within---or near---a regime where gravitational waves can merge the binary within the age of the universe.
Therefore, the confirmed detection of a sub-parsec SMBHB would be invaluable for understanding the relevant physics that enable SMBHB mergers and drive SMBH growth.

Possibly the best current indication that SMBHBs do merge is the strong evidence for a stochastic gravitational wave background recently reported by Pulsar Timing Array experiments \citep{iPTA2022, NANOG-GWB:2023, EPTA-GWB:2023, ParkesPTA-GWB:2023, CPTA-GWB:2023, Miles+25}.
Such a background was long predicted from a cosmological population of in-spiraling SMBHBs \citep[\eg][]{Carr:1980, Phinney:2001, SesanaVecchio+2008}.
The observation significance will grow steadily with time, and we may eventually identify the loudest single sources as continuous waves above the background \citep[\eg][]{Gardiner+25, Becsy_indivPTA+2025}.
The GWs from the final coalescence of SMBHBs are also the target for future space based GW detectors like LISA \citep{LISA_LRR_2023, LISA:RedBook:2024} and TianQin \citep{TianQin}, or proposed techniques to probe micro-Hz frequencies \citep[\eg][]{ZwickUOP+2025, Foster_muHz_binres+2025, Caliskan_GaiaGW+2024}.
Each of these GW measurements encodes some information about the physics that promoted these inspirals and mergers (\eg from waveform parameters like mass, spin, and phase), but the bulk of the data about the astrophysical conditions that manufactured their coalescence require electromagnetic identification of their host galaxy.

Numerous methods have been pursued for identifying SMBHBs electromagnetically, such as finding variability in source lightcurves at the binary orbital period (or its harmonics) \citep{Haiman+2009, DDLens:2018}; searching for offset, variable, or double peaked broad emission lines \citep[\eg][]{Gaskel:BLR:1996, Ju+2013, Runnoe+2015, Runnoe+2017, Runnoe+2025, Kelley_BLR:2021, MohammedBog+2025}; looking for X-ray excesses from circumbinary gas dynamics \citep{Roedig_SEDsigs+2014, Farris:2015:Cool}; identifying periodicity in continuum reverberations
\citep[\eg][]{Bogdanovic+2008, ShenLoeb:2010, 
PG1302MNRAS:2015a,
DHLighthouse+2017, NguyenBogIII:2020, Ji_CBD_BLR+2021, DottiPol+2022, Bertassi+2025, Malewicz+2025}; or connecting helicity, kinks, or X-shapes in jets and radio structures to some underlying binary orbital motion \citep[\eg][]{Roos:1988, Roos:1993, Romero+2000, MerrittEker:2002, Kun+2015:PG1302, Britzen_JetWiggles+2023} (see the review by \citealt{DOrazioCharisi:2023} and the references therein for a thorough discussion).
Nearly all of these techniques, however, require that the binary be active, similar to a single SMBH in an active galactic nucleus (AGN).
While galaxy mergers are generally believed to stimulate AGN activity in the post-merger system \citep[\eg][]{Comerford_AGNFuel+2024}, because of the lack of constraints on the timescales between galaxy merger and SMBHB merger, it remains unclear if these epochs coincide with binary inspiral.
Moreover, the nearest (redshift $ \lesssim 0.5$) and most massive binaries (total mass $\gtrsim 10^9 \Msun$) like those most likely sourcing the gravitational wave background (GWB), most probably reside in massive, gas poor, elliptical galaxies as quiescent, inactive sources \citep[][but see also \citealt{Zhou:2025}]{Izquierdo-Villalba:2023,TruantSesana:CW+EM:2025,CellaTaylorKelley:2025, Veronesi:2025}. 
Therefore, discovering these systems electromagnetically require search techniques that do not necessitate large gas reservoirs and AGN-like feeding activity.

The circum-nuclear region around quiescent or low-luminosity SMBHs (like SgrA$^\star$ or M87$^\star$) are not completely devoid of gas, but rather, are typically modeled with a radiatively inefficient, two-temperature, advection-dominated accretion flow characterized by very low gas densities and Eddington fractions \citep{NarayanYi:ADAF:1994, Moscibrodzka:SagA*M87*:2016, EHT:M87:2019, EHT:SagA:2022}.
Some effort has previously been made to explore the dynamics and emission properties of SMBHBs in such advection-dominated accretion flows (ADAFs) by simulating Bondi accretion onto binaries in the phases immediately preceding (and during) merger \citep{FarrisLiuShap:2010:Bondi, Bode:2010, Bode:2012}.
They determined that such systems can remain detectable out to modest redshifts ($z \lesssim 1$), can potentially source periodicity at the binary frequency from shocks very near the binary components, and might induce distinct rise-and-drop behavior in the pre- and post-merger luminosity.

Other so-called low-luminosity AGN (LLAGN), are well modeled with a mixed-component accretion flow whereby an outer AGN-like thin disk switches to an ADAF solution at some characteristic ``truncation'' radius \citep{SLE:1976, Narayan:1996, Esin+1997, Abramowicz:rtrans:1998, Nemmen+2006, Nemmen+2014} (although the physical reason for the transition is not fully understood; \eg \citealt{Meyer:DiskEvap:1994, Liu:coronae:1999, Honma:1996, Manmoto:2000, GuLu:2000}).
SMBHBs in mild-to-low accretion rate environments can also present with such mixed-component accretion flows, but with compositions potentially distinct from the single SMBH scenario.
In particular, the flow around each binary component and the full binary itself can in theory select any three-element-permutation of thin, AGN-like disks or hot, ADAF-like torii.
This fact was noted by \citet{PG1302Nature:2015b} and \citet{KelleyHaiman:2019} in predicting optical variability characteristics from SMBHBs, 
and in \citet{Koudmani_unifiedAccModels:2024}, which explored multi-component subgrid models for single and binary BH accretion in galaxy scale simulations.
However, a detailed analysis of the available mixed-component binary accretion states and full spectral models of these configurations have not yet been pursued.

In this paper we develop a model for characterizing the full spectral energy distribution (SED) of these mixed-component binary accretion flows.
We describe the basic physical picture in Section~\ref{S:physical-picture}, the model for computing broadband SEDs in Section~\ref{S:spectral-characteristics}, and possible variability signatures in Section~\ref{S:photometric-signatures}.
We discuss how these models can be applied to search for low-luminosity SMBHBs, how their characteristics might clash with other known SMBH properties, and present a proof of concept application to a well-known SMBHB candidate PG1302-102 in Section~\ref{S:discussion}.
We summarize and comment on future directions in Section~\ref{S:conclusions}.

%
\section{Accretion states} \label{S:physical-picture}

There exist four general solutions for accretion flows onto black holes.
Which---for a given black hole mass $M$, viscous parametrization $\alpha$, and radius $r$ (in Schwarzschild radii $r_s$)---is generally set by the accretion rate $\Mdot$.
The standard thin disk solutions \citep{SS73, LBPringle:1974} occur for systems accreting below their Eddington limit $\Mdot_\edd$. These disks are Keplerian, optically thick, and quickly radiate viscously dissipated energy from the disk faces before it can be transported radially.
For accretion rates that exceed the system's Eddington limit, standard thin solutions remain valid down to a characteristic radius at which the local photon diffusion time exceeds the accretion time.
Interior to this ``trapping radius'' solutions are described as slim disks \citep{Katz:1977, Begelman:1979, Abramowicz:1988}.
Both of these solutions are \emph{cold} in the sense that their gas temperatures are small compared to the virial temperature.

Formally, thin disk solutions are valid for all sub-Eddington accretion rates provided the disk remains optically thick to absorption under blackbody cooling.
This optical-depth transition, however, occurs at exceptionally small accretion rates for SMBHs, so we do not consider it further \citep{SS73, NarayanYi:ADAF:1995}.\footnote{Formally, the ``middle'' and ``inner'' regions of standard thin disks become optically thin at modest $\dot M$, but this is never actualized because the electron-scattering-to-free-free opacity-transition-radius shrinks more rapidly than the absorptive optical depth with decreasing $\Mdot$. Thus, the bounding accretion rate is that computed from the ``outer'' region \citep[c.f.][]{Haiman+2009}.}
Below some critical rate $\Mdot_\crit \equiv \dot m_\crit \Mdot_\edd$, however, the flow has additional access to a class of \emph{hot}, optically thin solutions described by a two-temperature plasma.
If the flow remains radiatively efficient and geometrically thin, it is described by the Shapiro-Lightman-Eardley solution (SLE; \citealt{SLE:1976}).
SLE solutions, however, are thermally unstable and are expected to deteriorate into geometrically thick, radiatively inefficient, advection-dominated accretion flows (ADAFs) where viscously dissipated energy is retained by accreting gas.

The critical accretion rate $\dot m_\crit$, below which hot flows are available, can be estimated from heating and cooling balance. Consider the illustrative example of an initially optically thin accretion flow that is evaluated at large enough radius, $ \gtrsim r_\ie$, for it to be treated with a single temperature, $T$. We take $r_\ie \equiv 10^3 r_s$ \citep[\eg][]{NMQ:ADAF:1998}.
In such a regime, the dominant emission is free-free Bremsstrahlung with cooling rate $q_b^- \propto n_e^2 T^{1/2} \rightarrow (\alpha M)^{-2} \dot M^2 r^{-7/2}$, where we have expressed the electron number density, $n_e$, and $T$ through the ADAF self-similar solutions \citep{NarayanYi:ADAF:1994}.
When $q_b^-$ exceeds the self-similar viscous heating rate $q^+ \propto M^{-2} \Mdot r^{-4}$, the system will cool until it becomes optically thick, and transitions into a standard thin disk.
But for accretion rates below $\dot M_\crit \propto \alpha^2 r^{-1/2}$, the cooling cannot keep pace with viscous dissipation, the flow retains this energy as heat, and remains an ADAF.

More detailed numerical treatment of this calculation including more generalized cooling show that at large radius $\Mdot_\crit$ drops more steeply with radius, approximately $\propto r^{-3/2}$ \citep{Menou_rtr+1999}, but at small radii in the two-temperature regime, $r \lesssim r_\ie$, the critical accretion rate is approximately constant $\dot m_\crit \simeq (0.1 - 0.3)\alpha^2$ \citep{Esin+1997}.
We note that in the limit that $\alpha \lesssim 0.3$, there exists a second $\dot m_\crit^\prime > \dot m_\crit$ where formally $q_b^- > q^+$, but the gas remains hot because of compressional heating as material flows inward.
For our purposes, we consider a fiducial value $\alpha = 0.3$ so that $\dot m_\crit \approx \dot m_\crit^\prime$.
Lastly, there exists a third critical rate $\dot m_{\crit,\, e} \approx 10^{-3} \alpha^2$ below which electrons no longer cool effectively, but these systems are very dim and not considered \citep{XieYuan:mdot_crits:2012}.

%
\begin{figure}[t]
    \centering
    \includegraphics{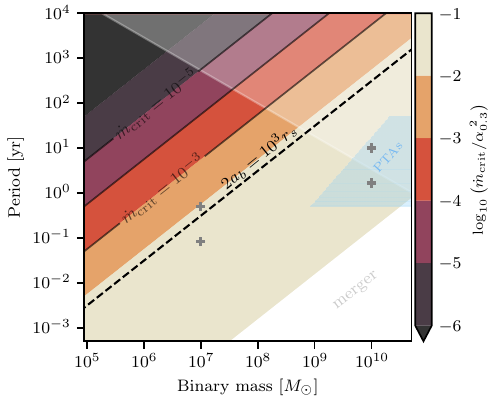}
    \caption{
    Critical accretion rates evaluated at the inner edge of a circumbinary disk ($r=2a_b$) for binaries with mass $M$ and period $P_b$ (and viscosity $\alpha = 0.3\alpha_{0.3}$).
    Below the indicated $\dot m_\crit$, hot advection-dominated solutions are available.
    Binaries that would contribute to the GWB at redshift $z=0.3$ are highlighted in blue. 
    Grey pluses represent fiducial binary systems used throughout.
    }
    \label{fig:dmcrits}
\end{figure}
%

%
\subsection{Cold or hot accretion}

Below  $\dot m_\crit$, when both hot and cold solutions are feasible, which is chosen and under which conditions is not well understood.
Observational data, however, suggests that whenever hot solutions are accessible, they are selected (the so-called ``strong-ADAF principle''; \citealt{NarayanYi:ADAF:1995}).
We will operate under this assumption from here-out, and comment to what extent this is an appropriate choice later on.

If the accreting object is a binary of total mass $M$ and semi-major axis $a_b$, the outer circumbinary (CBD) flow may be hot if the accretion rate is below $\dot m_\crit(2a_b)$.
We evaluate the critical rate at $2a_b$ because this generally corresponds to the last annulus of stable, non-intersecting orbits in the restricted three body problem \citep{RudPac:1981} and coincides with the evacuated cavity observed in hydrodynamic simulations \citep[\eg][]{AL94, MacFadyen:2008, D'Orazio:CBDTrans:2016, Ragusa_cavs+2020, MaheshMcW+2024}.
For radii beyond $r_\ie$, we take $\dot m_\crit = \dot m_\adv (r/r_\ie)^{-3/2}$, and interior to this $\dot m_\crit = \dot m_\adv = 0.2 \alpha^2$. 
Figure~\ref{fig:dmcrits} illustrates contours of $\dot m_\crit(2a_b) / \alpha_{0.3}^2$ over the space of binary masses and periods $P_b$ in Figure~\ref{fig:dmcrits} for $\alpha = 0.3 \alpha_{0.3}$.
The color indicates the accretion rate below which the inner edge of the CBD is susceptible to becoming an ADAF.
The beige regions correspond to $\dot m_\crit \simeq \dot m_\adv$, and the progressively darker zones to the radial suppression of $\dot m_\crit$ in the single-temperature disk.
The dashed black line denotes those binaries for which $2 a_b = r_\ie$.
Systems above this line have single-temperature CBDs and are progressively less susceptible to becoming ADAFs with increasing period (decreasing mass).
Below this line, the inner regions of the CBD are most prone to becoming advection-dominated with $\dot m_\crit = \dot m_\adv$.
The light-blue region shows those that would be detectable by PTAs at redshift $z=0.3$.
The grey crosses represent fiducial systems referenced in the following analyses.
We see that the  majority of binaries with year-or-shorter periods---and particularly any binary contributing to the GWB in PTAs---will accrete from a hot circumbinary flow if the outer feeding rate is below a modest accretion rate of $\dot m_\adv \simeq 10^{-1.7} \alpha_{0.3}^2$.

For reference we also include as the transparent white region the radius at which standard thin disks become gravitationally unstable for $\alpha = 0.3$ and $\dot m = \dot m_\adv$ (\citealt{Haiman+2009}, and note \citealt{Haiman2009:Erratum}).
Larger accretion rates and lower values of $\alpha$ expand this region, and vice versa.
Systems in this region are either accreting above $\dot m_\crit$ and are self-gravitating,\footnote{Though the inclusion of physics beyond standard alpha disks (\eg magnetic fields, energy injection from star formation) can restore Toomre stability.} or are accreting below their critical rate and are an ADAF.

%
\begin{figure*}[t]
    \centering
    \includegraphics[width=0.8\textwidth]{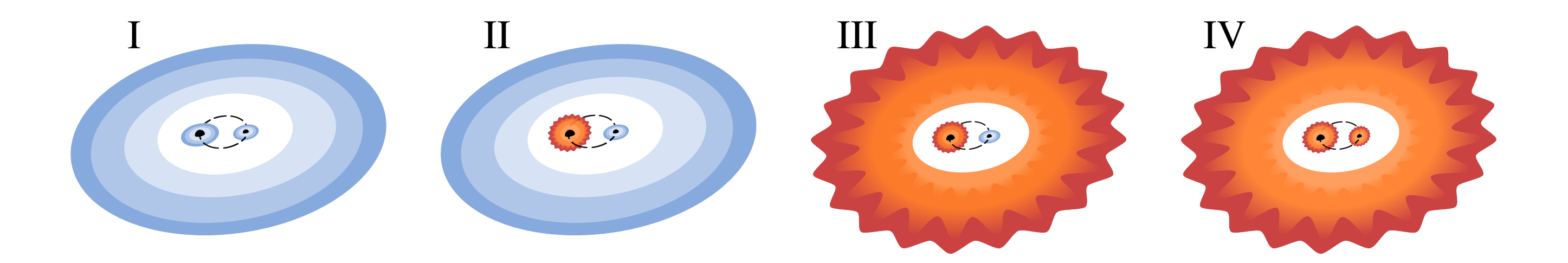}
    \includegraphics{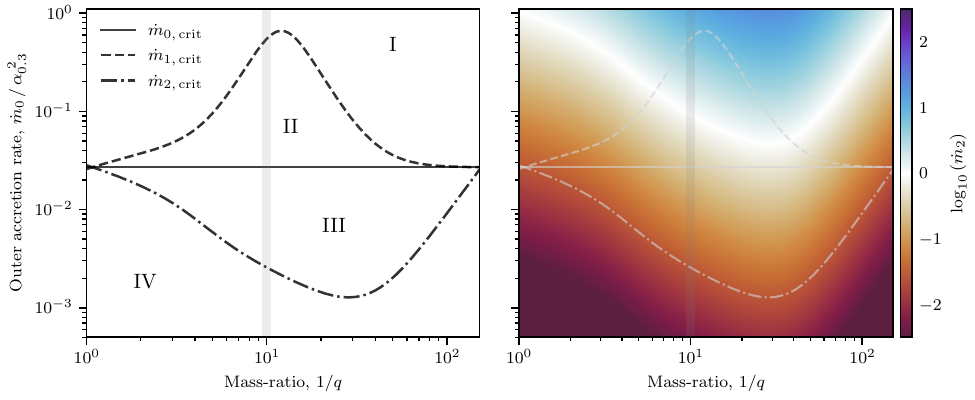}
    \caption{Characterization of multi-component accretion states. (\emph{Top}) Schematics illustrating available combinations of thin disk (blue) and ADAF (red) components.
    (\emph{Bottom-left}) Critical accretion rates for the full binary ($\dot m_{0, \crit}$; solid), primary ($\dot m_{1, \crit}$; dashed), and secondary ($\dot m_{2, \crit}$; dot-dashed) as a function of the binary mass ratio $1/q$ and mass supply rate $\dot m_0$.
    The Roman numerals indicate the regions of $\dot m_0 - q$ space corresponding to different accretion geometries as illustrated in the schematics above.
    The vertical grey band shows our fiducial choice of mass ratio for Section~\ref{S:spectral-characteristics}.
    (\emph{Right}) The colormap shows the secondary accretion rate $\dot m_2$, and the thin white band indicates when the secondary starts accreting above its Eddington limit (top, blue regions) under the assumed form of $\lambda(q)$.
    $\dot m_2$ only ever mildly exceeds its Eddington limit, such that this effect is not significant for our analysis (see text).
    }
    \label{fig:regions}
\end{figure*}
%

%
\subsection{Multi-component flows}

Generally, accretion onto a binary of mass ratio $q\leq1$ surrounded by a CBD is characterized by three accretion rates: that in the circumbinary disk , $\dot M_0$, set at scales much larger than the binary semi-major axis; that onto the primary component of the binary, $\dot M_1$; and that onto the secondary component, $\dot M_2$.
In Eddington units, let
\begin{align}
    \dot m_0 \equiv \frac{\Mdot_0}{\Mdot_\edd(M)}\,,\; \text{and}\quad \dot m_i \equiv \frac{\Mdot_i}{\Mdot_\edd(M_i)}
 \label{eq:eddrates}
\end{align}
for each component mass $M_1 = M / (1 + q)$ and $M_2 = q M / (1 + q)$.
We generally assume that mass is conserved, \eg $\dot M_0 = \dot M_1 + \dot M_2$, but this is generally not the case, especially in the very low-$\dot m$ limit. 
We discuss how including outflows may alter our results in Appendix~\ref{A:parameter-variation}.

How the outer accretion rate $\Mdot_0$ is distributed between each of the binary components is non-trivial and is typically determined numerically in hydrodynamics simulations.
High resolution results generally suggest that the secondary over accretes relative to the primary, such that $q$ evolves towards unity \citep{Farris:2014, Duffell:2020, DittmannRyan:2024} (however, see also \citealt{YoungBairdClarke:2015}).
The accretion splitting is described by the function $\lambda(q) \equiv \Mdot_2 / \Mdot_1$.
As a fiducial prescription we adopt
\begin{align}
    \lambda(q) = q^{-0.25} \, e^{-0.1/q} + \frac{50}{ (12 q)^{3.5} + (12 q)^{-3.5} }
 \label{eq:lambda}
\end{align}
provided by \citet{KelleyHaiman:2019} (from data originally generated in \citealt{Farris:2014}).

Given $\Mdot_0$ and conservation of mass, $\lambda(q)$ uniquely defines all three Eddington-normalized accretion rates $\dot m_j$ (Equation~\ref{eq:eddrates}; $j \in \{0,1,2\}$).
Under the strong ADAF principle, if any $\dot m_j$ fall below $\dot m_\crit$, the associated flow will become hot, optically thin, and advection-dominated.
In the left panel of Figure~\ref{fig:regions} we illustrate these critical rates $\dot m_{j,\,\crit}$, assuming $2a_b < r_\ie$ (and evaluated for $\alpha_{0.3}$), for increasingly unequal mass binaries (growing $1 / q$) at supply rate $\dot m_0$. 
The critical rate separating hot and cold solutions for the circumbinary disk is simply $\dot m_\adv$ shown by the solid, horizontal line.
The form of $\lambda$ in Equation~\eqref{eq:lambda} determines the critical curves for the primary (dashed line) and the secondary (dash-dotted line).
Because the secondary BH generally over-accretes relative to the primary, $\lambda(q)$ separates the space of binary mass ratios and feeding rates into four distinct accretion regions. 
Regions I and IV correspond to single-component accretion flows where all three disks are cold (I) or hot (IV).
However, for intermediate mass ratios and mildly sub-Eddington outer accretion rates, there exist two distinct multi-component accretion flows:
In Region II, the circumbinary and secondary disks are cold and thin, but the primary is feeding below the critical rate and transitions to a hot flow.
Conversely, in Region III, the circumbinary and primary disks are hot, but the secondary remains above the critical rate and so must be cold and thin.
Here we comment that because the capture of gas into orbit around the primary/secondary component generally involves tidal compression and shocks \citep{ShiKrolik:2015, Tiede:2022}, in Region II in particular, the transition from cold CBD material to a hot circum-primary torus is well justified because material will be hot and puffed upon capture.

One possible complexity is that at some points in $\dot m_0 - q$ space the secondary starts accreting above its Eddington limit.
The right panel of Figure~\ref{fig:regions} shows the secondary accretion rate $\dot m_2$ over the considered range of $q$ and $\dot m_0$.
The red-orange regions (lower portion) are sub-Eddington, while the blue areas (upper) are super-Eddington, with the white band designating $\Mdot_2 = \Mdot_\edd(M_2)$.
These accretion flows may transition from thin to slim disk solutions if the trapping radius becomes sufficiently large.
\citet{Watarai:SuperEdd:2006} provides an expression for the trapping radius (their Equation 28) in terms of only the accretion rate, $r_{\rm trap} = (0.25\dot m) \, r_s$.
Thus, for $r_{\rm trap}$ to exceed the inner-most stable circular orbit (ISCO; for a Schwarzschild black hole), $\dot m$ must exceed $\sim 10$.
This is almost never the case for $\dot m_2$ in Figure~\ref{fig:regions} -- with the exception of $q=0.3$ at $\dot m_0 \sim 1$, where $\dot m_2 \approx 20$. 
Even in this case, the effect would be to slightly reduce the effective temperature of the inner-most annuli, truncating the high-energy portion of the cold blackbody spectra.
This would have no effect on our conclusions, so we treat all solutions in Figure~\ref{fig:regions} above $\dot m_\crit$ as standard thin disks and reserve full treatment of super-Eddington binary accretion to future work.

%
\section{Spectral characteristics} \label{S:spectral-characteristics}

An accreting binary of given mass ratio and outer accretion rate uniquely specifies an accretion region in Figure~\ref{fig:regions}.
Each region consists of a different combination of hot and cold accretion components, and would thus be expected to present with different spectral characteristics.
To model the SEDs of each region, we treat the full system as a linear combination of three separate accretion flows: one for the outer, circumbinary disk, and one for each circum-single minidisk (see Section~\ref{S:discussion} for discussion on this assumption).
We refer to this as the three-disk, or binary-ADAF (BADAF), model.
We assume that the CBD extends from an inner edge at $2a_b$ out to $100a_b$ and that the circum-single disks subsist from the Schwarzschild ISCO of the component BH, to the tidal truncation radius, $R / a_b = 0.27 q^{\pm 0.3}$ from the binary companion.
The plus corresponds to the secondary component and the minus to the primary \citep{Paczynski:1977, Roedig_SEDsigs+2014}.

For the cold solutions we model a given disk as an optically thick $\alpha$-disk. 
The dominant cooling mechanism is thermal blackbody emission, so that the spectrum is characterized by the effective temperature profile 
\begin{align}
    T_i^4(r) = \frac{GM_i\dot M_i}{8\pi \sigma r^3} \left[ 1 - \left( \frac{r_{s, \,i}}{r} \right)^{1/2} \right]
\end{align}
with $r_{s, \,i}$ the Schwarzschild radius of mass $M_i$ and $\sigma$ the Stefan-Boltzmann constant.
The luminosity per frequency $\nu$ is then
\begin{align}
    L_{\nu, i} = 2\pi \int_{r_{\rm in, i}}^{r_{\rm out, i}} \mathcal{B}_\nu \left[ T_i(r) \right] r dr 
\end{align}
evaluated at the relevant inner- and outer-radii and with $\mathcal{B}_\nu[T]$ Planck's function.

The hot solutions, in contrast, are optically thin and radiation emerges from the combined cooling of electrons through Bremsstrahlung, synchrotron, and inverse Compton emission.
To determine the characteristic spectral features of a hot flow around a given black hole, we adopt the general formalism of \citet{Mahadevan:1997} through a modified implementation of \texttt{llagnsed} \citep{Pesce:LLAGNSED:2021}.
The general approach is to empirically find the electron temperature which balances the volume-integrated electron heating and cooling rates.
In order to do so, we assume that the electrons and ions are in thermal equilibrium at $10^3 r_s$ as discussed in Section~\ref{S:physical-picture}.
The energy balance equation for the electrons is
\begin{align}
    \delta\, Q^+ + Q^{\ie} = Q^{-} + Q^{\adv, e} \ .
 \label{eq:ebalance}
\end{align}
The left hand side represents the volumetric heating rate of electrons where $Q^+$ is the total rate of viscous dissipation and $\delta$ is the fraction directly deposited into electrons.
The viscous heating rate for ions, then, is $(1 - \delta)Q^+$, and $Q^\ie$ specifies the rate of energy transfer from ions to electrons through Coulomb collisions.
The value of $\delta$ is not well-constrained, but turbulent modeling suggests $\delta \approx 0.1 - 0.5$ \citep{Howes:2010, Rowan:deltae:2017, Kawazura:deltae:2019}.
%

%
\begin{figure*}[t!]
    \centering
    \includegraphics{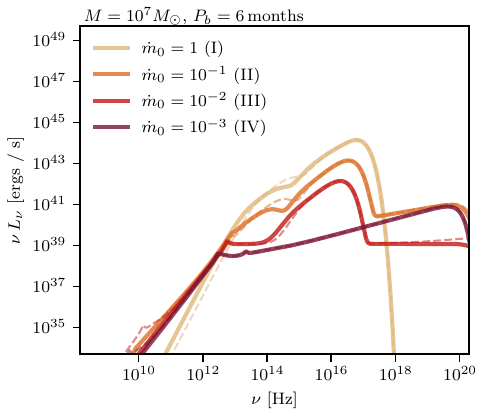}
    \includegraphics{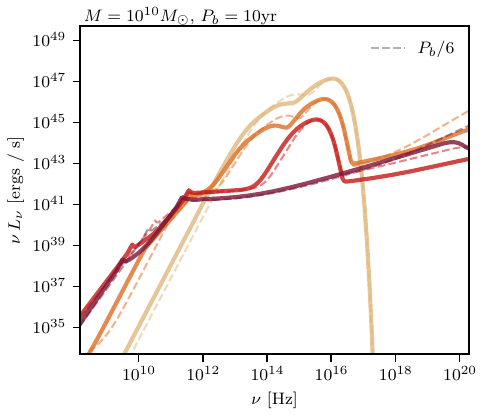}
    \caption{
    Full broadband SEDs for the fiducial binaries from our multi-component three-disk model:
    (\emph{Left}) $M = 10^7 \Msun$, $P_b = 6\,\unit{months}$, and 
    (\emph{Right}) $M = 10^{10} \Msun$, $P_b = 10\,\unit{years}$.
    Dashed lines are for a period a factor of six shorter (a month and $\sim 1.5$yr, respectively).
    }
    \label{fig:spectra_comp}
\end{figure*}
%

The right hand side of Equation~\eqref{eq:ebalance} comprises the total volumetric cooling rate, where the total radiated energy is
\begin{align}
    Q^- = \int_0^\infty d\nu \; \left(  L_\nu^\brems + L_\nu^\synch + L_\nu^\comp \right)
\end{align}
(see Appendix A in \citealt{Pesce:LLAGNSED:2021} for the particulars of computing each $L_\nu$).
In the ADAF solutions, the rate of heat advection is $f Q^+$ such that $f$ sets the fraction of the viscously dissipated energy that gets advected.
Most of this heat is advected with the ions, but under values of $\delta \gg m_e / m_p$, some amount, $Q^{\adv,e}$, is advected with electrons.

To compute spectral energy distributions (SEDs), we assume a fiducial hydrodynamic state with $\alpha = 0.3$, $\delta = 0.3$, and a ratio of gas pressure to magnetic pressure, $\beta = 10$.
Formally, the value of $f$ depends on both radius and accretion rate, but for advection-dominated solutions it's close to unity, so we adopt $f=1$.
For simplicity, we also nominally assume that there are no outflows so that $\Mdot_0 = \Mdot_1 + \Mdot_2 = {\rm const}$.
However, we note that outflows have been argued as a generic feature of ADAF solutions because of their positive Bernoulli numbers (\citealt{BB:ADIOS:1999}; but see also \citealt{Quataert:Gruzinov:2000}), so
we explore the effect of including them through a power law profile $\dot M \propto r^s$ \citep{Stone:ADIOS:1999} in Appendix~\ref{A:parameter-variation} (as well as that of varying $\delta$, $\beta$, and $f$).
\footnote{Outflows may also be present in configurations where the secondary accretes above its Eddington limit \citep[e.g.,][]{Kitaki:2018}, but this effect may be weak since the secondary is only ever mildly super-Eddington in the considered setups (e.g., Figure~\ref{fig:regions}).}

For a binary with mass ratio $q=0.1$ we compute the resulting SEDs from our three-disk model for four outer accretion rates corresponding to each of the regions in Figure~\ref{fig:regions}.
These SEDs are shown in Figure~\ref{fig:spectra_comp} for two characteristic binaries: one with total mass $10^7 \Msun$ and a period of $P_b=6$ months (left), and another with $M = 10^{10} \Msun$ and $P_b = 10$ years (right).
The former is representative of a binary that might be present in modern optical time-domain surveys like Rubin's LSST \citep{KelleyHaiman:2019, KelleyLens+2021} or Roman's High-Latitude Time-Domain Survey \citep{RomanPLCs+2023}, and the latter of those expected to most strongly contribute to the GWB in PTAs \citep{NANOG-SMBHBs:2023}. 

One can see that in Region I the spectrum is expectedly thermal, with a blue/UV-notch because of the lack of emission from the assumed circumbinary cavity (\citealt{Roedig_SEDsigs+2014}; although simulation studies have shown this to likely be spurious, \eg \citealt{Farris:2015:Cool})
In Region II, when the flow around the primary becomes hot, the "notch" remains due to overlapping CBD and secondary thin disk spectra, but the spectrum gains both a radio component from the associated hot primary synchrotron emission as well as a component in the hard X-rays (and $\gamma$-rays) primarily from the free-free emission.
Region III is characterized by similar synchrotron and Bremsstrahlung wings that generally match the full-hot SED of Region IV, but with a blue, thermal bump from the cold disk around the secondary BH. 
We note that the small peaks produced in the sub-millimeter and radio portions of Regions III and IV correspond to the synchrotron peak frequency. They are generally sourced near the solutions inner-most annuli such that the lower-energy peak coincides with the inner-edge of the CBD and the higher-frequency one to the inner circum-primary flow, potentially signaling the binary accretion geometry.

The dashed lines show the SEDs of equivalent binary accretion flows with orbital period decreased by a factor of six.
The main effect of changing the binary period is to move the location of the spectral notch in regions I \& II.
The hot components of the flow are minimally dependent on the binary period because in the considered examples, the radius where ions and electrons thermally equilibrate is larger than the binary semi-major axis (see Figure~\ref{fig:dmcrits}), and most of the hot emission is dominated by annuli near the binary components. 

\begin{figure*}[t!]
    \centering
    \includegraphics{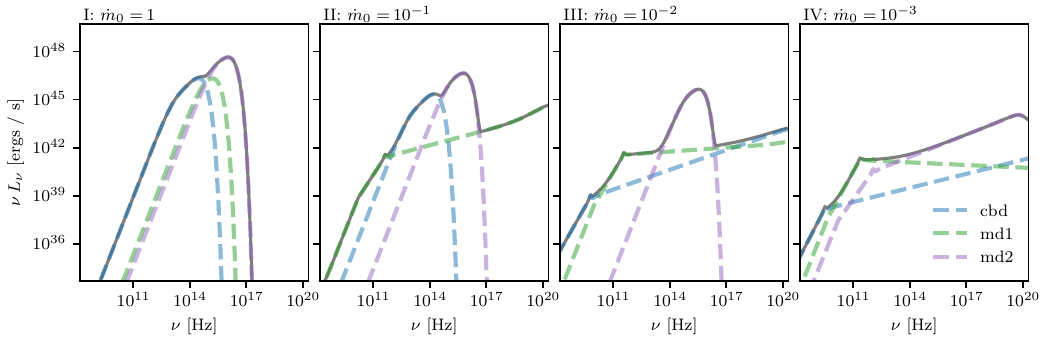}
    \caption{SED decomposition into contributions from the circumbinary disk (cbd), the primary minidisk (md1), and the secondary minidisk (md2) for the binary with $M = 10^{10} \Msun$ and $P_b = 10\unit{yr}$. 
    The solid grey line shows the total SED.
    The accretion rate and associated Region are indicated in the top left of each panel.
    }
    \label{fig:spectra_decomp}
\end{figure*}

To better illustrate the relative contribution of each disk component, we also show in Figure~\ref{fig:spectra_decomp} the SEDs for the $M = 10^{10} \Msun$ binary with $P_B = 10\unit{yr}$ decomposed into that from the circumbinary disk (cbd), the primary (md1), and the secondary (md2).
This decomposition is representative of all the SEDs considered here.
The peak emission is always sourced from the secondary disk because of it's comparatively small ISCO and large relative accretion rate $\dot m_2$ under preferential accretion $\lambda$, while the source of radio and X-ray wings varies with $\dot m_0$.
In Region II, the wings naturally are produced by the hot circum-primary flow. 
In Region III the lowest and highest energy emission is from the CBD, but the shoulder at millimeter wavelengths is sourced by the primary.
Of note, in Region IV all emission from the optical to $\gamma$-ray emerges from the circum-secondary flow, and the comparatively large UV$-$X-ray slope from accretion onto the smaller secondary component might distinguish the binary from a standard hot solution around a single-BH of equal total mass

%
%

%
\section{Photometric signatures} \label{S:photometric-signatures}

A primary method for identifying SMBHB candidates in electromagnetic surveys is to search for periodicity at the binary orbital period, $P_b$ (or its harmonics).
For cold thin solutions, when emission variability is approximated through modifications to the accretion rate, the timescale for these changes to be reflected in the radiation is approximately $(\alpha \Omega_k)^{-1}$, the local thermal time (with $\Omega_k$ the Keplerian orbital frequency)---or that required for the new heating rate to alter the effective temperature of the disk.
In order to clearly see modulations at $P_b$, this must be faster than the binary orbital period.
This is generally true in the minidisks, so we would expect emission from either of the circum-single components to imprint the binary orbital motion.
However, this would imply that emission from the CBD will not vary at $P_b$ if the modulation only affects the dissipation rate in the mid-plane.
Hydrodynamics simulations, though, show that the rotating binary potential strongly perturbs the CBD structure through shocks and density waves.
If these perturbations deposit primarily in the mid-plane (or deep layers) of the CBD, the above argument holds; 
but if they effect a significant portion of the vertical column, the relevant timescale becomes that required to communicate the perturbation to the emission layers, $H / c_s \sim \Omega_k^{-1}$ (for scale height $H$ and sound speed $c_s$).
Under this assumption, the binary can source periodic variability in regions where $P_b \gtrsim \Omega^{-1}_k$, which includes both the minidisks and the inner-regions of the circumbinary disk (which appears to be the case in recent radiation-magnetohydrodynamic simulations; \citealt{Tiwari:bRMHD:2025, Chan_RMHD_mds+2025}).

Conversely, for hot solutions the flow is optically thin and any hydrodynamic perturbation to the steady state can potentially source variability at $P_b$.
However, if the perturbation period is shorter than the associated cooling time, the response at $P_b$ will be damped, de-phased, and possibly washed out of the emission.
To estimate which portions of our SED may vary at $P_b$, we compute frequency averaged cooling times for each mechanism assuming the perturbations act on the ion and electron fluids similarly.
A relativistic distribution of electrons with characteristic energy $\langle E_e \rangle \equiv \langle \gamma \rangle m_e c^2$ will cool over a time $n_e \langle E_e \rangle / q^-$ for a given emission power density $q^-$.
We assume a Maxwell-J\"{u}ttner distribution of electrons such that the mean Lorentz factor is
\begin{align}
    \langle \gamma \rangle = \int_1^\infty \gamma n(\gamma) d\gamma
 \label{eq:gamma-avg}
\end{align}
with distribution function
\begin{align*}
    n(\gamma)d\gamma = \frac{\gamma \sqrt{\gamma^2 - 1}}{\theta_e K_2(1/\theta_e)} \,\exp{\left(-\frac{\gamma}{\theta_e} \right)} d\gamma 
\end{align*}
given in terms of the dimensionless temperature $\theta_e \equiv k_b T_e / m_e c^2$.
$K_2$ is the Modified Bessel function of the second kind and the distribution function is normalized such that $\int_1^\infty n(\gamma)d\gamma = 1$.

For thermal Bremsstrahlung, the emission density in a neutral plasma is
\begin{align}
    q^-_\brems = \Lambda_{ff} \,\bar g_B \, n_e^2\, T_e^{1/2} \,\Phi(T_e)
\end{align}
where
\begin{align*}
    \Lambda_{ff} = \frac{2^5 \pi q_e^6}{3 h m_e c^3} \sqrt{\frac{2\pi k_b}{3 m_e}} \ ,
\end{align*}
we choose a Gaunt factor $\bar g_B = 1.2$, and $\Phi(T_e) = 1 + 4.4 \times 10^{-10} T_e$ is a relativistic correction \citep{Rybicki:RadiativeProcesses}.
The characteristic cooling time associated to free-free emission is then
\begin{align}
    t_\brems = \frac{\langle \gamma \rangle m_e c^2}{\Lambda_{ff} \bar g_B \Phi(T_e)} n_e^{-1} T_e^{-1/2}.
 \label{eq:tbrems}
\end{align}
The power density in synchrotron and inverse Compton emission can be calculated as
\begin{align}
    q^-_{\rm \{synch,\ comp\}} = \frac{4}{3}\, n_e \sigma_T c\, u_{\rm \{mag,\ rad\}} \int_1^\infty (\gamma^2 - 1) n(\gamma) \, d\gamma
 \label{eq:tcool}
\end{align}
where $u_{\rm x}$ is the energy density in either magnetic fields, $u_{\rm mag} = B^2 / 8\pi$, or incident radiation, $u_{\rm rad} = L^{\rm synch} / (4 \pi r^2 c)$ where we've assumed the bath of low-energy photons are from the synchrotron component. 
We compute each of these frequency averaged cooling times as a function of radius by evaluating Equations~(\ref{eq:gamma-avg}-\ref{eq:tcool}) numerically with the radial profiles $n_e(r)$ and $B(r)$ given by the ADAF self-similar solutions (for fiducial values of $\alpha$, $\beta$, $\delta, f)$.
The electron temperature profile and incident radiation energy density are extracted from our solutions that balance total volumetric heating and cooling around a central source of mass $M$ and Eddington fraction $\dot m$, as described in Section~\ref{S:spectral-characteristics}.

\begin{figure}[t!]
    \centering
    \includegraphics{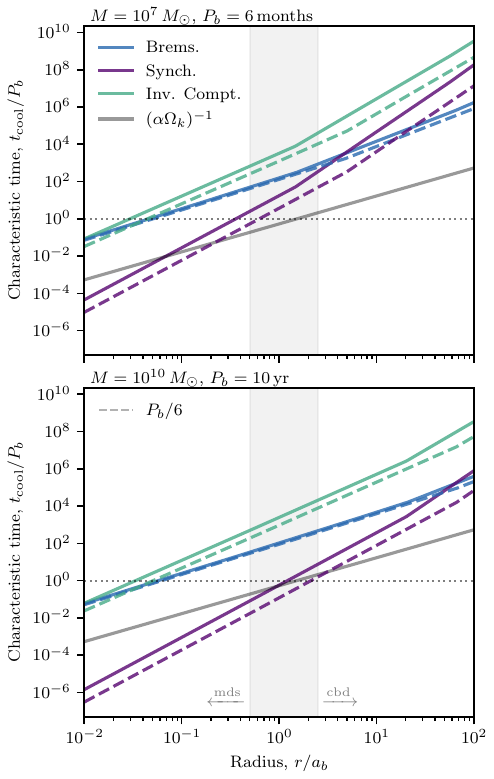}
    \caption{Characteristic cooling profiles for Bremsstrahlung, synchrotron, and inverse Compton emission from the ADAF self-similar solutions for a $10^7 \Msun$ binary with a six month period (upper panel) and a $10^{10} \Msun$ binary with a ten year period (lower panel) as a function of radius in units of binary semi-major axes. 
    Both panels are for an outer mass supply rate $5 \times 10^{-3} \dot{M}_\edd$.
    Dashed lines show an equivalent binary with one-sixth the period.
    The grey region shows typical scales of the circumbinary cavity such that regions to its left are generally part of the circum-single flows (mds) and to the right part of the circumbinary flow (cbd).
    The solid grey line illustrates the characteristic inflow time for hot, advection-dominated solutions, which we regard as a lower limit (see text).
    }
    \label{fig:cooling-profiles}
\end{figure}

Radial profiles of these grey cooling times are illustrated in Figure~\ref{fig:cooling-profiles} for the reference binaries from Section~\ref{S:spectral-characteristics} in units of the binary orbital period---with the outer accretion rate now fixed to $\dot m_0 = 5 \times 10^{-3}$.
Dashed lines once again denote a reference binary with one-sixth the orbital period (one month and a year-and-a-half respectively).
The solid grey line indicates the characteristic inflow time, $(\alpha \Omega_k)^{-1}$, at which a typical fluid element will be advected into the central source for the considered hot solutions.\footnote{We note that another statement of the hot, advection-dominated regime is when the characteristic inflow time of an annulus is equal to or faster than the standard thermal time, $(\alpha \Omega_k)^{-1}$.
Electrons can cool faster than the local thermal time in the ADAF solutions so long as the ions remain hot (which can be very roughly inferred from the free-free cooling time).
}
However, radial advection in the strong tidal regime (\eg in the circumbinary cavity) may be slower, so we regard this as a lower limit.
To guide the interpretation, the light grey band illustrates the typical radii of the circumbinary cavity, such that the regions leftwards of the band generally belong to the circum-single disks (mds) and those rightwards represent the circumbinary flow (cbd). 
One can see that for each system, the Bremsstrahlung and inverse Compton cooling times are significantly longer than the binary perturbation period for all radii except the inner-most regions of the flow, and are generally longer than the typical inflow time.
Binary perturbations would not only be damped in the associated emission, but the perturbed fluid elements would also typically get accreted. 
The synchrotron cooling, however, is generally faster than the binary period (and comparable to or less than the inflow time) in the circum-single flows, and can thus vary at $P_b$. 
The synchrotron emission will vary most strongly near the peak frequency (which is also generated from the inner-most annuli), and so the sub-mm peaks from the circum-primary flow in Regions II-IV (see, \eg Figure~\ref{fig:spectra_decomp}) would vary at $P_b$.
The more massive reference system ($10^{10} \Msun$) can even feasibly source variability at the binary period into the CBD at $\sim \textrm{few} \times  a_b$, which would likely manifest as modulations to the lower synchrotron peak and into the radio tails.

%
%

%
\begin{figure*}[t!]
    \centering
    \includegraphics{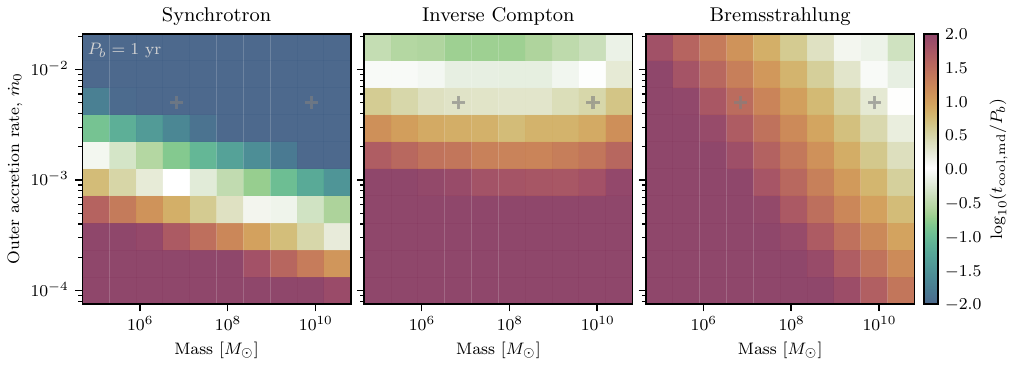}
    \caption{Characteristic cooling times evaluated at the typical truncation radius for an equal-mass binary ($r \simeq 0.3 a$) for a one-year-period binary with indicated mass and outer accretion rate. Tiles right-wards and up-wards of the white tiles (with green/blue hues) have characteristic cooling times shorter than the binary orbital period and could source periodic variability around a binary component. Those left-wards and down-wards (with orange/red hues) radiate too slowly to source distinct periodic emission at $P_b$.
    Grey crosses indicate the tiles associated to the models in Figures~\ref{fig:spectra_comp}~and~\ref{fig:cooling-profiles}.
    }
    \label{fig:cooling-periods}
\end{figure*}
%

The sensitivity of the synchrotron cooling time in particular can be understood by expressing the radial dependence of the ADAF magnetic flux density at fixed accretion rate in units of $a_b$, 
\begin{align} \nonumber
    B\big( r / a_b \big) \propto (1 + \beta)^{1/2} \dot m^{1/2} M^{1/3} P_b^{-5/6} \left( \frac{r}{a_b} \right)^{-5/4} \ ,
\end{align}
such that, up to changes in the solution temperature profile, $t_{\rm synch} / P_b  \sim n_a \dot m^{-1}$ where $n_a \equiv a_b / r_s$ is the binary separation in gravitational radii. 
In Figure~\ref{fig:cooling-profiles}, the smaller binary has $n_a \simeq 700\, (200)$ and the bigger one $n_a \simeq 50\, (15)$.
Thus, the more compact and near-merger systems might imprint the binary periodicity in synchrotron emission at scales comparable to the binary itself.

To better visualize how these effects vary mutually with $M$ and $\dot m$, we plot in Figure~\ref{fig:cooling-periods} the characteristic cooling times at a typical minidisk truncation radius $0.3 a_b$ for each mechanism across a grid of central masses and accretion rates with fixed binary period $P_b = 1\unit{yr}$.
Green/blue regions (top and right of each panel) indicate cooling times that are shorter than one year, and orange/red regions where the typical cooling times are longer than the binary period.
The grey crosses show the systems considered in Figures~\ref{fig:spectra_comp}~and~\ref{fig:cooling-profiles}.
As noted, synchrotron emission can generally source variability at $P_b \sim 1$yr for $\dot m \gtrsim 10^{-3}$ and can do so for even lower rates at larger masses.
Inverse Compton emission can only source distinct variability at one year when the disk density is large, at accretion rates near the critical rate $\dot m \sim 10^{-2}$. 
Bremsstrahlung emission can only potentially reflect binary periodicities of order years at the largest binary masses $M \gtrsim 10^{10} \Msun$ and near critical accretion rates.

Therefore, for nearly all possible binary masses and accretion rates, systems accreting from mixed-component flows would typically show hydrodynamic variability at the binary orbital period only in their thermal blackbody radiation from any thin disk components (\eg in the optical/UV around the secondary component or infrared from the CBD); as well as in the low-frequency synchrotron emission provided it is not otherwise dominated by a jet (this possibility is discussed in more detail in Section~\ref{S:discussion}). 
We also note that in addition to this hydrodynamic variability, emission at all frequencies from the circum-single flows should vary due to Doppler boosting or gravitational lensing.

%
\section{Discussion} \label{S:discussion}
%

%
\subsection{Implications for electromagnetic SMBHB searches}

Much work has been invested in seeking spectral signatures of compact (sub-pc separation) binary accretion ranging from continuum spectral notches \citep{GultekinMiller_SEDGaps:2012, Roedig_SEDsigs+2014, GenerozovHaiman:2014} to unique or variable broad emission profiles \citep{NguyenBogI:2016, Kelley_BLR:2021} to unusual radio or jet morphologies \citep{MerrittEker:2002, Britzen_JetWiggles+2023} to abnormal X-ray spectral indices, X-ray-to-optical flux ratios, or Fe-K$\alpha$ lines \citep{McKFeZoltan:2013, Geoff:2017, dAscoli+2018, Malewicz+2025}.
Despite these efforts, associated spectroscopic searches for binarity on sub-pc separations have been generally unsuccessful \citep[\eg][]{Saade:2020, Saade:2024, Guo+2020} and many suggested features are equally available to standard, single-BH AGN \citep{LobanovZensus:2001, Farris:2015:Cool, LiuErac:2016, Charisi+2016, NguyenBogII:2019}.
Similar to these previous works, our model for multi-component binary accretion does not provide a smoking-gun signature for a central SMBHB.
It does, however, offer a number of possible hints, and importantly, a step towards a holistic, multi-wavelength analysis of binary candidates that links binary signatures across the electromagnetic spectrum. 

Given an outer accretion rate and a binary mass ratio, our formalism predicts binary-AGN spectral shapes from radio to gamma-ray frequencies, and the wavelengths where periodicity might arise or be suppressed.
This offers a number of opportunities for vetting accreting binary candidates. 
For example, consider the case of a binary candidate discovered in an AGN through periodic photometric variability. 
The timescale of oscillations provides an estimate for the putative binary orbital period. Given a central BH mass measurement (\eg from broad emission lines) and the average in-band flux, bolometric corrections can be used to estimate the Eddington-normalized accretion rate onto the binary. 
Through the character of the variability, one can additionally constrain the binary mass ratio \citep[\eg][]{DHM:2013:MNRAS, Farris:2014, PG1302Nature:2015b}. From such estimates for the accretion rate and binary parameters, one can use the results presented here to estimate the broadband SED of the multi-component accreting system and take first steps towards a number of consistency checks and predictions. 
We discuss some of these possibilities here.

\paragraph{Binary spectral fitting}
Applying bolometric corrections derived from broad classes of AGN to estimate the total accretion rate may not be valid if the luminosity in band is derived from accretion onto, say, the secondary component of a binary.
A fitting procedure, similar to fitting thin-to-ADAF transition radii in single-BH multi-component flows \citep[\eg][]{Nemmen+2014}, could be carried out for binaries over outer accretion rates and mass ratios that determine the mix of accretion phases. 
Multi-wavelength observations could then probe consistency across the spectrum. 
Furthermore, radio loud systems generally require an additional jet component to the spectrum, and adding this to the SED can provide independent constraints on the jet inclination, disentangle which parts of the spectrum are dominated by the accretion flow or the jet, and provide vital information for vetting multiple concurrent binary signatures.

\paragraph{Multi-wavelength variability} 
Given the expected SED, one can check if the band for which periodic variability is observed is consistent with expectations from cooling-time arguments in Section~\ref{S:photometric-signatures}. Multiple parts of the spectrum could be probed in the time domain to test this variability expectation. 
Interestingly, the analysis in Section~\ref{S:photometric-signatures} suggests that in addition to standard optical-to-X-ray variability from thermal disk emission, periodic accretion variability is expected at radio to sub-mm frequencies. 
This provides interesting possibilities for radio variability searches, \citep[\eg][]{OVRO+2011, ONeillPKS_2131:2022, OVROcand2:2025},
although we note that modeling the radio emission of standard LLAGN typically requires a jet component or additional non-thermal electrons that may wash out this contribution \citep{Quataert:llagn:1999, Ozel:2000, LiuWu:2013}.
As noted, periodic Doppler boosting can also cause variability in parts of the SED that are dominated by emission from material bound to either of the individual components.
Combining jet- and disk-spectra also allows a consideration of where to apply disk-variability vs. jet-variability models
\citep{DOrazioCharisi:2023}.
A final interesting possibility in the case of outflows is that the winds from hot accretion components (e.g. from the primary and CBD) might interact periodically to create another source of variability, but we leave characterizing this possibility for future work.

\paragraph{VLBI tracking possibilities}
These SEDs provide millimeter-wavelength predictions for accreting SMBHBs which can be used to estimate observability with Very Long Baseline Interferometry (VLBI), as well as corresponding possibilities for directly resolving binary or circumbinary disk structures \citep[\eg][]{DOrazioLoebVLBI:2018,Kovacevic_PLC+astrometric_SMBHB:2024, Zhao_VLBI_SMBHB+2024}.

\paragraph{Reverberation}
The derived SED can be used to predict the properties of reverberated emission that depends on the incident continuum.
This includes periodic dust echos \citep{DHLighthouse+2017}, CBD irradiation by minidisk emission \citep{Bang_CBDIrrad+2025}, polarization from scattered minidisk emission \citep{DottiPol+2022}, broad line variability \citep{NguyenBogI:2016, Bertassi+2025}, and variable X-ray reflection spectra \citep{McKFeZoltan:2013, Malewicz+2025}.
Reverberation features depend on the incident emission from the triple-disk system and have thus far been primarily modeled assuming thin disk emission or associated coronal features. Including multi-component illumination allows further exploration of reverberated features and offers a consistent/alternative way to include incident high-energy emission. Importantly, the reverberated emission can create its own, additional features in the SED---like an infrared bump from a dusty torus---while also generating variable continuum and spectral features that must be self-consistently modeled with the underlying SED.

\paragraph{Spectral shape}
While the shape of the spectrum could become a powerful predictor of binarity and the geometry of the accretion flow, \eg through notches, X-ray spectral slopes, the frequency and multiplicity of synchrotron peaks, and other distinct features observed in Figure \ref{fig:spectra_comp}, we caution against their over interpretation. An accurate description of these features would require a more detailed treatment which departs from the linear, three-disk superposition approach we have taken in this first study.
The full binary multi-component flow must me modeled in order to capture binary-induced shock heating and accretion splitting in the case of non-trivial thermodynamics and geometries.

\vspace{5pt}
The above pieces must be consistent across the SED model.
To illustrate this, we provide a case study for the SMBHB candidate in the quasar PG 1302 below.

%
\subsection{Application to SMBHB candidate PG1302-102}

%
\begin{figure*}[t!]
    \centering
    \includegraphics{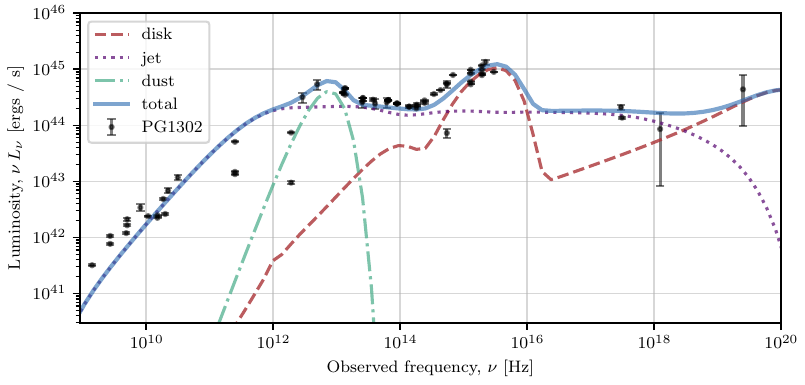}
    \caption{Case-study of the binary candidate PG1302-102 fit with our multi-component three-disk model (in red, dashed) in Region II to a binary of total mass $10^{9.4}\Msun$, mass ratio $0.3$, and rest-frame period $P_b = 4\unit{yr}$. 
    Observed data are shown as the black markers \citep{NED}.
    PG1302 is a blazar, so we include in purple (dotted) a simple jet model that we emphasize is primarily illustrative.
    In green (dot-dashed), we include a cold dust component to capture the infrared bump.
    The full SED is shown in blue (solid) and captures the predominant features of the system including the thermal blue bump and high-energy emission.
    }
    \label{fig:pg1302-model}
\end{figure*}
%

A particularly intriguing SMBHB candidate is the blazar PG1302-102 (hereafter PG1302; \citealt{Graham+2015a}).
PG1302 exhibits correlated multi-wavelength variability in the UV and optical with a rest-frame period of $\sim 4 \unit{yrs}$ \citep{Charisi+2018, Xin:PG1302:2020}.
It remains uncertain, though, if the periodicity is persistent and not simply a few-cycle manifestation of standard quasar variability \citep{LiuGezari+2018, ZhuThrane:2020}.
The SMBHB model for the periodicity in PG1302 is the relativistic Doppler boosting of emission from a standard, radiatively efficient disk around the secondary black hole \citep{PG1302Nature:2015b}.
In this model, the binary has total mass $\approx 10^{9.4} \Msun$, mass ratio $q \lesssim 0.3$, and a primary component that is accreting below its critical rate.
Thus, the putative binary model for PG1302 exists in a mixed-component accretion flow of Region II.
As a proof of concept, in Figure~\ref{fig:pg1302-model} we plot the available broadband SED data for PG1302 alongside our mixed-component SED model with $M = 10^{9.4}$, $P_b=4$yr (rest frame), $\dot m_0 = 0.06$, $q=0.3$, $\alpha = 0.3$, $\delta =0.3$, $s=0.5$, and redshift $z = 0.278$.
The secondary disk accurately characterizes the optical/UV ``blue bump'', as expected, and the hot, advection-dominated flow around the primary accounts for the observed hard X-ray and $\gamma$-ray fluxes (although, by itself would under-predict the soft X-rays, perhaps related to the well-known soft-excess in observed AGN spectra; \citealt{Singh_softX+1985, Arnaud_softX:1985}).

The model is not particularly accurate at low frequencies, however, because PG1302 is a blazar and launches an extended radio jet.
We include in Figure~\ref{fig:pg1302-model} a \citet{BlandfordKonigl:1979} model for the jet following the formalism of \citet{Spada:jets:2001, YuanCuiNarayan:2005} with jet half-opening angle $0.1$, bulk Lorentz factor $10$, energy density of accelerated electrons $\epsilon_e = 0.06$, and amplified magnetic field energy density $\epsilon_b = 0.05$.
We assume that the jet is launched from the hot flow around the primary and that the jet feeding rate is $0.1\%$ of the primary's accretion rate.
We take the observer angle to be $10^\circ$, but note that the jet model is meant to be illustrative as opposed to a detailed quantitative analysis of the PG1302 jet \citep[c.f.,][which derives a few degree inclination angle within the binary model]{Kun+2015:PG1302}.
We also include as the green, dashed-dotted curve a cold dust feature as a simple blackbody with covering factor 0.3 and characteristic temperature $120 \unit{K}$ illuminated by the integrated optical and UV emission of the base binary SED model (red, dashed) and placed at a distance required to match the SED. 
We note that this addition is simply demonstrative and a more careful approach is required to discern dust properties. Furthermore, such a feature might equally be produced from the irradiation of the CBD itself by the thermal circum-secondary emission \citep[\eg][]{LeeOkazakiHayasaki:2024, Bang_CBDIrrad+2025}.

While we do not attempt to constrain the nature of PG1302 here,
we conclude that the full broadband SED is consistent with that of the predicted BADAF and jet model. 
Furthermore, the required binary parameters, jet inclination, and dust radius/geometry needed to match the PG 1302 SED could introduce
constraints on the Doppler boost model \citep{PG1302Nature:2015b} and its reverberated emission \citep{DHLighthouse+2017, DottiPol+2022, Bertassi+2025} in combination with binary models of the jet morphology and kinematics \citep{Kun+2015:PG1302}. 
Furthermore, we learn that the Doppler-boost variability ought to arise primarily from the optical and UV emission associated to the secondary's cold, thermal disk.
Doppler boosted emission from the orbital motion of the primary could also arise for the radio-sub-mm and X-ray jet emission, and this should interestingly follow a different variability amplitude scaling with spectral index and orbital inclination.
\citep[\eg][]{PG1302Nature:2015b, ONeillPKS_2131:2022, DOrazioCharisi:2023}
While we leave further analysis to a future study, we note that this more holistic picture of the accreting binary SED opens up new and potentially powerful possibilities for vetting SMBHB candidate systems.

\subsection{Ramifications for binary orbital decay}

The characteristic timescale for binary orbital decay is generally set by the time required for the binary to interact with mass of order itself,
\begin{align}
    t_{m} \approx 4.5 \times 10^7 \,\eta_{0.1}\, \dot m_0^{-1} \, \unit{yr} \ ,
 \label{eq:salpeter}
\end{align}
where the coefficient is the Eddington limited mass doubling time ($t_s$) at a fiducial radiative efficiency $\eta = 0.1 \eta_{0.1}$.
When the system is being fed above its critical rate $\dot m_0 > \dot m_\crit$, the outer disk is thin and binary inspiral is mediated by periodic, non-linear tidal stripping of material from the CBD's inner edge \citep{MacFadyen:2008, ShiKrolik:2015, Tiede:2022}.
The relative strength of this interaction and the efficiency with which it delivers mass to the binary components can modify $t_m$ and feasibly reduce the inspiral time by an order of magnitude or two \citep{Dittmann:2022,Tiede:2025}.
In the high accretion limit, however, it is also possible for thin CBDs to induce binary expansion \citep{MirandaLai+2017, Munoz:2019}.
While this represents a relatively small set of parameter space for mass ratios $\gtrsim 0.1$ \citep[see, \eg][]{Tiede:2020, Zrake+2021, DOrazioDuffell:2021, Franchini_eccSG+2024}, for $q \lesssim 0.05$ disk-mediated binary outspiral appears more general \citep{Dempsey:2021, DittmannRyan:2024}.

On the other hand, if the system is feeding below the critical rate $\dot m_0 < \dot m_\crit$ and selects a hot, advection-dominated mode (due to either the initial conditions at large radii or significant shock heating sourced by the periodic binary potential), the relevant process is slightly different.
Namely, the hot solutions are significantly sub-Keplerian and have radial inflow rates that are a substantive fraction of the Keplerian orbital velocity (see, \eg Section~\ref{S:photometric-signatures}).
As a result the binary components (and their associated accretion flows) are evolving through the ``bath'' of the surrounding flow and slowly shrinking their orbit via a process akin to gas dynamical friction (\citealt{Ostriker:99}; although this remains an approximation because the flow retains semi-Keplerian motion and likely resonances). 
\citet{Narayan:adafDF:2000} computed the associated linear drag from the advection-dominated self-similar solutions
\begin{align}
    t_{m} = \frac{\alpha}{\dot m_0} \frac{q^{-1}}{\ln{(q^{-1})}}\, t_s 
    %
    \simeq 2.9 \times 10^9\, \eta_{0.1}\, (\dot m_0 / \dot m_\adv)^{-1} \,\unit{yr} \ ,
 \nonumber
\end{align}
longer than Equation~\eqref{eq:salpeter} by a factor of $\dot m_\adv / \dot m_0$ for fiducial parameters $\alpha = 0.3$ and $q = 0.1$.
One can see that this characteristic inspiral time from hydrodynamic drag, though, is also inversely proportional to $q$, so smaller binaries will take commensurately longer to merge.
While this estimate may hold in the small-$q$ limit, for large mass ratios, the solution is assuredly non-linear, and so may meaningfully deviate from the estimated evolution.
The precise rate of orbital decay would require dedicated simulation, but absent extremely efficient angular momentum exchange, must generally correspond to Equation~\eqref{eq:salpeter}.

In both scenarios, the accretion flow more readily facilitates binary decay for larger mass ratios $q \gtrsim 0.05$.
Assuming for a moment that gas interaction is a necessary component for creating compact SMBHBs that merge through GW emission, this may suggest that more equal mass systems are more easily brought to merger. 
Systems with significant mass discrepancies may first need to grow their mass ratio to $\sim 0.1$ before they can emit meaningful gravitational radiation.
This also corresponds with the mass ratios most likely associated to loud gravitational wave emitters either as continuous wave sources in the PTA band or as the highest signal-to-noise sources in LISA.

%
\subsection{Caveats and future directions}

Our model offers a tractable framework for studying mixed-component and low luminosity SMBHBs, but 
could in the future be improved a number of ways.
Foremost, we have relied heavily on a specific form for the accretion splitting $\lambda(q)$ based on purely hydrodynamical simulations of thin disk accretion.
While the general behavior of $\lambda(q)$ appears robust in this setting (\eg that the secondary over-accretes relative to the primary), it has some sensitivity to system parameters \citep[\eg][]{Duffell:2020, DittmannRyan:2024} and may not hold into the regime of hotter, thicker solutions \citep{YoungClarke:2015, YoungBairdClarke:2015}.
However, it remains unclear how the splitting behaves when the circumbinary flow transitions to a hot, optically thin mode as opposed to just a hotter radiatively efficient one.

Moreover, we have enforced the conservation of mass (and noted that this may be a poor approximation for advection dominated solutions), but we have not explicitly treated the conservation of angular momentum.
Namely, the tidal coupling between the binary and the accretion flow will source an angular momentum current that may additionally alter the profiles and emission features.
A more complete treatment would also account for this transfer of angular momentum (as well as how it couples to the binary orbit evolution).
We have also only treated the case of binaries accreting below their Eddington limit, but future extensions ought to address when the system is fed at rates exceeding the Eddington condition.
This would augment existing configurations with an additional set that include slim disk solutions (as discussed briefly in Section~\ref{S:physical-picture}).
We also note that while we have focused on systems where the accretion flow is dominated by gas pressure, evidence from modeling M87$^\star$ and SgrA$^\star$ \citep{EHT:M87:Bfield:2021} suggest that such radiatively inefficient flows may have magnetically arrested disks (particularly at the lowest accretion rates considered; \citealt{Narayan:MAD:2003, Tchekhovskoy:MAD:2011}), and that such strong magnetic fields may be a generic feature of ADAF-like environments.
Recently, it has also been noted that binary accretion can also manifest in such magnetically arrested configurations \citep{MostWang:2024, WangMostHopkins:BMAD:2025}.
Exploring such scenarios for hot, advection dominated binary accretion constitutes worthwhile follow-up work.

Formal modeling of LLAGN, as noted, typically involves a solution that transitions from a thin disk at large radii to a hot, advection-dominated flow at small radii. 
In our model, the circumbinary flow could theoretically also consist of such a thin disk at large radii before transitioning to a hot flow near its inner edge (\eg in Region III or IV).
The associated SED in Region III would likely appear similar to that in Region II because of the outer thermal disk, and in Region IV the SED would likely resemble Region III with the thermal bump shifted to lower frequencies (akin to the ``big red bump'' of some LLAGN; \citealt{Lasota:1996, Quataert:llagn:1999}).
For binaries with larger orbital periods where $\dot m_\crit$ begins to fall with radius (see, \eg Figure~\ref{fig:dmcrits}), there also exists the possibility not only that the CBD might comprise of both an outer-thin and inner-hot solution, but that each of the circum-single disks might also possess this structure.
Whether or not this is realized, though, would likely be sensitive to the specific physics of how material is captured into the circum-single flows.
To this end, we have also assumed that systems accreting below $\dot m_\crit$ always become hot and advection-dominated (the strong-ADAF principle).
We posit that this is a good assumption for the circum-single flows because capture from the CBD inner-edge generally involves strong compression and shocks that would initialize such flows as hot. However, this is less clear for the CBD because at large distances the solution must become thin\footnote{Unless the feeding rate is very low---below those generally considered here---and initially hot \citep[\eg the feeding of SgrA$^\star$ from colliding stellar winds;][]{Melai:1992, Cuadra:2006}.} (see Section~\ref{S:physical-picture}), and under what conditions such systems transition remains an open question \citep[\eg][]{Liska:GRMHD:2022}. 

We have, additionally, focused tacitly on circular binaries only. 
However, numerical studies of accreting eccentric binaries have exhibited ``accretion switching'' between binary components whereby the dominant accretor can alternate between the primary and secondary \citep{Dunhill+2015, Zrake+2021, Siwek:2023}.
This effect arises from the relative precession between the eccentric binary and the eccentric CBD, and generally operates on timescales of hundreds$-$thousands of binary orbits \citep[][but the ``switch'' itself can occur in $\sim$ few$-$tens of orbits]{ThunKley:2017, MunozLithwick:2020,  Grcic:2025}.
In some regimes, this behavior could cause the circum-single components of the binary accretion flow to alternate between cold and hot solutions and potentially lead to observable spectral changes for short period (weeks$-$months) binaries.

Lastly, our core assumption is that the spectral features of such SMBHBs are the linear super-position of three separate accretion flows, but a more detailed treatment should self-consistently link the three solutions in a full non-linear hydrodynamic solution. A first step to this would likely be simulating such binaries in a two temperature accretion flow with some form of the relevant heating and cooling (see, \eg \citealt{Sadowski:2TRMHD:2017, Chael:2tempRMHD:2025}).

%
\section{Conclusions} \label{S:conclusions}

We have developed a model for the electromagnetic properties of circumbinary accretion flows in the low-$\dot M$ limit like might be expected for SMBHBs in the local universe or for the massive binaries contributing most strongly to the nano-Hz GWB.
In particular, we have considered the cases where the feeding rates fall below a critical value and gain access to an alternate class of advection-dominated solutions.
Following some preliminary previous work \citep{PG1302Nature:2015b, KelleyHaiman:2019, Koudmani_unifiedAccModels:2024}, we have identified four possible combinations of standard, cold and hot, advection-dominated accretion flows as a function of the system feeding rate. 
This includes the possibility of two unique mixed-component accretion flows where part of the solution is hot and part cold.
We developed a three-disk model for generating full broadband SEDs for each of these four accretion regions as a function of the mass supply rate.
These systems, and the multi-component modes in particular, are generally characterized by a big-blue bump from cold accretion onto the secondary and radio, X-ray wings from hot accretion onto the primary.
These properties, while generally consistent with typical AGN-like accretion flows plus coronae around single-BHs, 
provide a framework for multi-wavelength analyses of SMBHB candidates linking observations across the electromagnetic spectrum.

Additionally, within the framework of our model, we performed a variability analysis to estimate which portions of the binary SEDs might fluctuate at the binary orbital frequency.
We determined that the cold, blackbody components sourcing the thermal-blue bump in the optical/UV are most likely to reflect hydrodynamic variability at the binary period.
By comparing the characteristic cooling times to the binary orbital frequency in the ADAF components, we argued that only the low-frequency synchrotron emission might also echo the orbital period hydrodynamically (unless the mass supply rate is very near $\dot m_\crit$ or the binary mass exceeds $\sim 10^{10} \Msun$). 
We comment that all emission from annuli bound to one of the binary components ought to also vary at the binary period from Doppler or lensing effects, but that the nature of this effect will depend on the relative significance of the emission against other components (\eg the CBD or jet) and on the binary inclination angle.

We have also estimated the characteristic timescales for binary orbital decay and determined that these are generally comparable and sufficient to instantiate a merger within a Hubble time regardless of the accretion configuration.
Interestingly, from compiling previous results, we comment that such orbital decay appears generally inefficient for binary mass ratios $\lesssim 0.05$.
This suggests that those binaries that transition into the sub-parsec (and GW emitting) regimes might preferentially be those with comparable masses, but a more detailed numerical analysis would be required to confirm this for the mixed-component solutions.

Lastly, we have applied our model to the SMBHB candidate, PG1302-102, which would exist in Region II of our analysis.
We found that our model can consistently account for the system's optical and high energy emission, but that the low-frequency component---because it is a blazar---is indistinguishable against the systems jet emission. 
Nonetheless, we have demonstrated that mixed-component and purely advection-dominated accretion onto SMBHBs can consistently account for many documented properties of AGN.
More detailed modeling will be necessary to identify properties unique to binary systems as the community pursues a first confirmation of a sub-parsec SMBHB and potential multi-messenger source.

%
\begin{acknowledgements}
The authors thank Zoltan Haiman, Kimitake Hayasaki, Roman Gold, Ramesh Narayan, Martin Pessah, and Lorenz Zwick for enlightening discussions on this topic over the years.
The authors also graciously thank the anonymous referee for their constructive comments.
This work was supported by the European Union’s Horizon 2023 research and innovation program under Marie Sklodowska-Curie grant agreement No. 101148364, by Sapere Aude Starting grant No. 121587 through the Danish Independent Research Fund, and by the National Science Foundation through Astronomy and Astrophysics Grant No. 2511544.
\end{acknowledgements}
%

%
\appendix
\renewcommand\theequation{A\arabic{equation}}
\renewcommand\thefigure{A\arabic{figure}}
\setcounter{equation}{0}
\setcounter{figure}{0}

%
\vspace{-20pt}
\begin{figure*}[h!]
    \centering
    \includegraphics{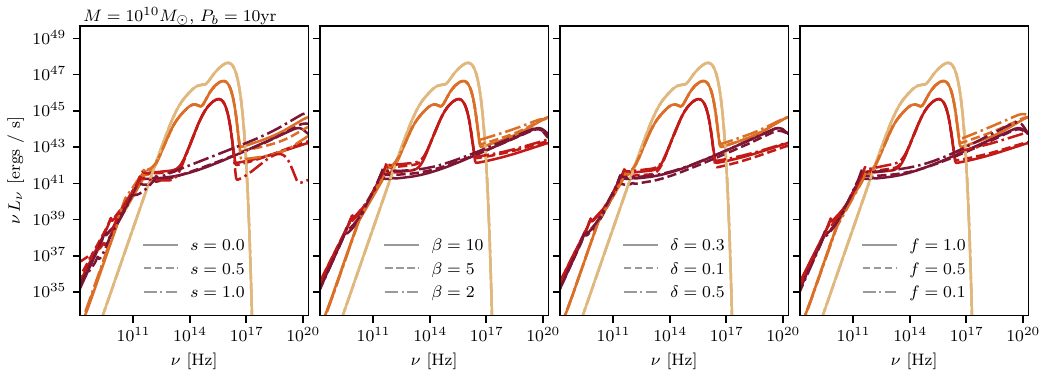}
    \caption{Sensitivity of our model SEDs to intrinsic parameters $s$, $\beta$, $\delta$, and $f$.
    The colors correspond to the same accretion rates in Figure~\ref{fig:spectra_comp}.
    }
 \label{fig:param-variation}
\end{figure*}
%

%
\section{Dependence on model parameters}

\label{A:parameter-variation}

We document the sensitivity of the SEDs generated in Section~\ref{S:spectral-characteristics} to our intrinsic model parameters in Figure~\ref{fig:param-variation}; specifically, the mass loss rate $s$ (left panel; where formally $\dot m(r) = \dot m_0 (r / r_{\rm cav})^s$ and $r_{\rm cav} = 2a_b$), the relative strength of magnetic fields $\beta$ (left-middle panel), the fraction of viscously dissipated energy deposited directly into electrons $\delta$ (right-middle panel), and the fraction of viscous heat that is advected $f$ (right panel).
The SEDs are not particularly sensitive to reasonable variation of $f$, $\delta$ or $\beta$ (in the regime that the magnetic energy density is sub-dominant to the fluid's thermal energy---we leave the magnetically dominated regime for future study), with only minor adjustments to the SEDs.
Variations under differing mass loss rates are more significant with the high-energy luminosities varying by an order of magnitude or two.

%
\section{Variability for short-period binaries}
\label{A:short-periods}

\begin{figure*}[h!]
    \centering
    \includegraphics{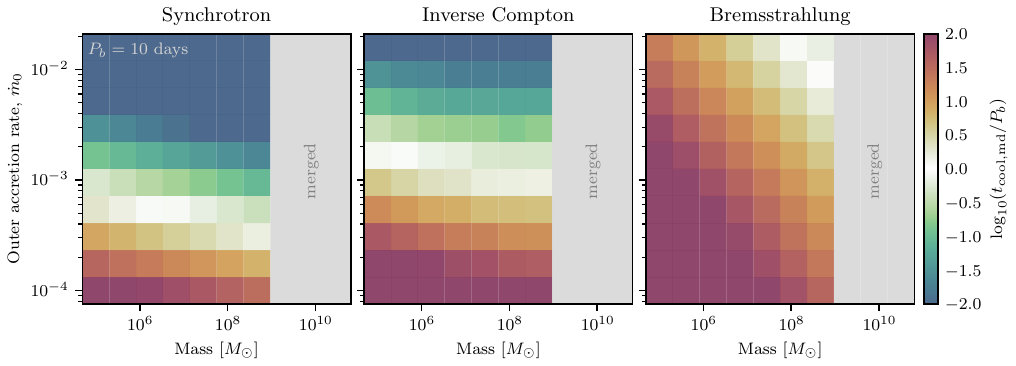}
    \caption{Same as Figure~\ref{fig:cooling-periods}, but for a reference period of 10 days. Systems in the grey region are disallowed because they would have merged.
    }
 \label{fig:cooling-10days}
\end{figure*}

Because the Rubin and Roman time-domain surveys will be sensitive down to binary periods of order days \citep{XinHaiman_LSSTshort:2021, RomanPLCs+2023}, we remake Figure~\ref{fig:cooling-periods} for a reference period $P_b = 10\,\unit{days}$ in Figure~\ref{fig:cooling-10days}.
We note that the large gas pressures and relative inflow rates of hot solutions imply that such binaries may retain their accretion flows all the way through merger \citep[\eg][]{Bode:2010, FarrisLiuShap:2010:Bondi}.

%
%

%
\bibliographystyle{mnras}
\bibliography{refs}

@ARTICLE{Chan_RMHD_mds+2025,
       author = {{Chan}, Chi-Ho and {Tiwari}, Vishal and {Bogdanovi{\'c}}, Tamara and {Jiang}, Yan-Fei and {Davis}, Shane W.},
        title = "{Radiative magnetohydrodynamics simulation of minidisks in equal-mass massive black hole binaries}",
      journal = {arXiv e-prints},
         year = 2025,
        month = may,
          eid = {arXiv:2505.02919},
        pages = {arXiv:2505.02919},
          doi = {10.48550/arXiv.2505.02919},
archivePrefix = {arXiv},
       eprint = {2505.02919},
 primaryClass = {astro-ph.HE},
       adsurl = {https://ui.adsabs.harvard.edu/abs/2025arXiv250502919C}
}

@ARTICLE{ThunKley:2017,
       author = {{Thun}, Daniel and {Kley}, Wilhelm and {Picogna}, Giovanni},
        title = "{Circumbinary discs: Numerical and physical behaviour}",
      journal = {\aap},
         year = 2017,
        month = aug,
       volume = {604},
          eid = {A102},
        pages = {A102},
          doi = {10.1051/0004-6361/201730666},
archivePrefix = {arXiv},
       eprint = {1704.08130},
 primaryClass = {astro-ph.EP},
       adsurl = {https://ui.adsabs.harvard.edu/abs/2017A&A...604A.102T}
}

@ARTICLE{MunozLithwick:2020,
       author = {{Mu{\~n}oz}, Diego J. and {Lithwick}, Yoram},
        title = "{Long-lived Eccentric Modes in Circumbinary Disks}",
      journal = {\apj},
         year = 2020,
        month = dec,
       volume = {905},
       number = {2},
          eid = {106},
        pages = {106},
          doi = {10.3847/1538-4357/abc74c},
archivePrefix = {arXiv},
       eprint = {2008.08085},
 primaryClass = {astro-ph.HE},
       adsurl = {https://ui.adsabs.harvard.edu/abs/2020ApJ...905..106M}
}

@ARTICLE{Guo+2020,
       author = {{Guo}, Hengxiao and {Liu}, Xin and {Zafar}, Tayyaba and {Liao}, Wei-Ting},
        title = "{Spectral energy distributions of candidate periodically variable quasars: testing the binary black hole hypothesis}",
      journal = {\mnras},
         year = 2020,
        month = feb,
       volume = {492},
       number = {2},
        pages = {2910-2923},
          doi = {10.1093/mnras/stz3566},
archivePrefix = {arXiv},
       eprint = {1907.06676},
 primaryClass = {astro-ph.GA},
       adsurl = {https://ui.adsabs.harvard.edu/abs/2020MNRAS.492.2910G}
}

@ARTICLE{Malewicz+2025,
       author = {{Malewicz}, Julie and {Ballantyne}, David R. and {Bogdanovi{\'c}}, Tamara and {Brenneman}, Laura and {Dauser}, Thomas},
        title = "{X-Ray Reflection Signatures of Supermassive Black Hole Binaries}",
      journal = {\apj},
         year = 2025,
        month = aug,
       volume = {989},
       number = {2},
          eid = {190},
        pages = {190},
          doi = {10.3847/1538-4357/adea75},
archivePrefix = {arXiv},
       eprint = {2504.14018},
 primaryClass = {astro-ph.HE},
       adsurl = {https://ui.adsabs.harvard.edu/abs/2025ApJ...989..190M}
}

@ARTICLE{Howes:2010,
       author = {{Howes}, G.~G.},
        title = "{A prescription for the turbulent heating of astrophysical plasmas}",
      journal = {\mnras},
         year = 2010,
        month = nov,
       volume = {409},
       number = {1},
        pages = {L104-L108},
          doi = {10.1111/j.1745-3933.2010.00958.x},
archivePrefix = {arXiv},
       eprint = {1009.4212},
 primaryClass = {astro-ph.HE},
       adsurl = {https://ui.adsabs.harvard.edu/abs/2010MNRAS.409L.104H}
}

@ARTICLE{OVROcand2:2025,
       author = {{de la Parra}, P.~V. and {Kiehlmann}, S. and {Mr{\'o}z}, P. and {Readhead}, A.~C.~S. and {Synani}, A. and {Begelman}, M.~C. and {Blandford}, R.~D. and {Ding}, Y. and {Harrison}, F. and {Liodakis}, I. and {Max-Moerbeck}, W. and {Pavlidou}, V. and {Reeves}, R. and {Vallisneri}, M. and {Aller}, M.~F. and {Graham}, M.~J. and {Hovatta}, T. and {Lawrence}, C.~R. and {Lazio}, T.~J.~W. and {Mahabal}, A.~A. and {Molina}, B. and {O'Neill}, S. and {Pearson}, T.~J. and {Ravi}, V. and {Tassis}, K. and {Zensus}, J.~A.},
        title = "{PKS J0805-0111: A Second Owens Valley Radio Observatory Blazar Showing Highly Significant Sinusoidal Radio Variability{\textemdash}The Tip of the Iceberg}",
      journal = {\apj},
         year = 2025,
        month = jul,
       volume = {987},
       number = {2},
          eid = {191},
        pages = {191},
          doi = {10.3847/1538-4357/addc60},
archivePrefix = {arXiv},
       eprint = {2408.02645},
 primaryClass = {astro-ph.HE},
       adsurl = {https://ui.adsabs.harvard.edu/abs/2025ApJ...987..191D}
}

@ARTICLE{OVRO+2011,
       author = {{Richards}, Joseph L. and {Max-Moerbeck}, Walter and {Pavlidou}, Vasiliki and {King}, Oliver G. and {Pearson}, Timothy J. and {Readhead}, Anthony C.~S. and {Reeves}, Rodrigo and {Shepherd}, Martin C. and {Stevenson}, Matthew A. and {Weintraub}, Lawrence C. and {Fuhrmann}, Lars and {Angelakis}, Emmanouil and {Zensus}, J. Anton and {Healey}, Stephen E. and {Romani}, Roger W. and {Shaw}, Michael S. and {Grainge}, Keith and {Birkinshaw}, Mark and {Lancaster}, Katy and {Worrall}, Diana M. and {Taylor}, Gregory B. and {Cotter}, Garret and {Bustos}, Ricardo},
        title = "{Blazars in the Fermi Era: The OVRO 40 m Telescope Monitoring Program}",
      journal = {\apjs},
         year = 2011,
        month = jun,
       volume = {194},
       number = {2},
          eid = {29},
        pages = {29},
          doi = {10.1088/0067-0049/194/2/29},
archivePrefix = {arXiv},
       eprint = {1011.3111},
 primaryClass = {astro-ph.CO},
       adsurl = {https://ui.adsabs.harvard.edu/abs/2011ApJS..194...29R}
}

@ARTICLE{Arnaud_softX:1985,
       author = {{Arnaud}, K.~A. and {Branduardi-Raymont}, G. and {Culhane}, J.~L. and {Fabian}, A.~C. and {Hazard}, C. and {McGlynn}, T.~A. and {Shafer}, R.~A. and {Tennant}, A.~F. and {Ward}, M.~J.},
        title = "{EXOSAT observations of a strong soft X-ray excess in MKN 841.}",
      journal = {\mnras},
         year = 1985,
        month = nov,
       volume = {217},
        pages = {105-113},
          doi = {10.1093/mnras/217.1.105},
       adsurl = {https://ui.adsabs.harvard.edu/abs/1985MNRAS.217..105A}
}

@ARTICLE{Singh_softX+1985,
       author = {{Singh}, K.~P. and {Garmire}, G.~P. and {Nousek}, J.},
        title = "{Observations of Soft X-Ray Spectra from a Seyfert 1 and a Narrow Emission-Line Galaxy}",
      journal = {\apj},
         year = 1985,
        month = oct,
       volume = {297},
        pages = {633},
          doi = {10.1086/163560},
       adsurl = {https://ui.adsabs.harvard.edu/abs/1985ApJ...297..633S}
}

@ARTICLE{Bang_CBDIrrad+2025,
       author = {{Bang}, Saemi and {Okazaki}, Atsuo T. and {Hayasaki}, Kimitake},
        title = "{Spectral Properties of Irradiated Circumbinary Disks around Binary Black Holes Governed by Hydrogen Opacities Dependent on Temperature and Density}",
      journal = {arXiv e-prints},
         year = 2025,
        month = jun,
          eid = {arXiv:2506.14141},
        pages = {arXiv:2506.14141},
          doi = {10.48550/arXiv.2506.14141},
archivePrefix = {arXiv},
       eprint = {2506.14141},
 primaryClass = {astro-ph.HE},
       adsurl = {https://ui.adsabs.harvard.edu/abs/2025arXiv250614141B}
}

@ARTICLE{Zhao_VLBI_SMBHB+2024,
       author = {{Zhao}, Shan-Shan and {Jiang}, Wu and {Lu}, Ru-Sen and {Huang}, Lei and {Shen}, Zhiqiang},
        title = "{How Many Supermassive Black Hole Binaries Are Detectable through Tracking Relative Motions by (Sub)millimeter Very Long Baseline Interferometry?}",
      journal = {\apj},
         year = 2024,
        month = jan,
       volume = {961},
       number = {1},
          eid = {20},
        pages = {20},
          doi = {10.3847/1538-4357/ad0da1},
archivePrefix = {arXiv},
       eprint = {2311.11589},
 primaryClass = {astro-ph.GA},
       adsurl = {https://ui.adsabs.harvard.edu/abs/2024ApJ...961...20Z}
}

@ARTICLE{Kovacevic_PLC+astrometric_SMBHB:2024,
       author = {{Kova{\v{c}}evi{\'c}}, Andjelka B. and {Songsheng}, Yu-Yang and {Wang}, Jian-Min and {Popovi{\'c}}, Luka {\v{C}}.},
        title = "{Bayesian Synthesis of Astrometric Wobble and Total Light Curves in Close Binary Supermassive Black Holes}",
      journal = {\apj},
         year = 2024,
        month = may,
       volume = {967},
       number = {1},
          eid = {30},
        pages = {30},
          doi = {10.3847/1538-4357/ad3729},
archivePrefix = {arXiv},
       eprint = {2404.17435},
 primaryClass = {astro-ph.IM},
       adsurl = {https://ui.adsabs.harvard.edu/abs/2024ApJ...967...30K}
}

@ARTICLE{ONeillPKS_2131:2022,
       author = {{O'Neill}, S. and {Kiehlmann}, S. and {Readhead}, A.~C.~S. and {Aller}, M.~F. and {Blandford}, R.~D. and {Liodakis}, I. and {Lister}, M.~L. and {Mr{\'o}z}, P. and {O'Dea}, C.~P. and {Pearson}, T.~J. and {Ravi}, V. and {Vallisneri}, M. and {Cleary}, K.~A. and {Graham}, M.~J. and {Grainge}, K.~J.~B. and {Hodges}, M.~W. and {Hovatta}, T. and {L{\"a}hteenm{\"a}ki}, A. and {Lamb}, J.~W. and {Lazio}, T.~J.~W. and {Max-Moerbeck}, W. and {Pavlidou}, V. and {Prince}, T.~A. and {Reeves}, R.~A. and {Tornikoski}, M. and {Vergara de la Parra}, P. and {Zensus}, J.~A.},
        title = "{The Unanticipated Phenomenology of the Blazar PKS 2131-021: A Unique Supermassive Black Hole Binary Candidate}",
      journal = {\apjl},
         year = 2022,
        month = feb,
       volume = {926},
       number = {2},
          eid = {L35},
        pages = {L35},
          doi = {10.3847/2041-8213/ac504b},
archivePrefix = {arXiv},
       eprint = {2111.02436},
 primaryClass = {astro-ph.HE},
       adsurl = {https://ui.adsabs.harvard.edu/abs/2022ApJ...926L..35O}
}

@ARTICLE{LiuGezari+2018,
       author = {{Liu}, Tingting and {Gezari}, Suvi and {Miller}, M. Coleman},
        title = "{Did ASAS-SN Kill the Supermassive Black Hole Binary Candidate PG1302-102?}",
      journal = {\apjl},
         year = 2018,
        month = may,
       volume = {859},
       number = {1},
          eid = {L12},
        pages = {L12},
          doi = {10.3847/2041-8213/aac2ed},
archivePrefix = {arXiv},
       eprint = {1803.05448},
 primaryClass = {astro-ph.HE},
       adsurl = {https://ui.adsabs.harvard.edu/abs/2018ApJ...859L..12L}
}

@ARTICLE{Begelman:1979,
       author = {{Begelman}, M.~C.},
        title = "{Can a spherically accreting black hole radiate very near the Eddington limit?}",
      journal = {\mnras},
         year = 1979,
        month = apr,
       volume = {187},
        pages = {237-251},
          doi = {10.1093/mnras/187.2.237},
       adsurl = {https://ui.adsabs.harvard.edu/abs/1979MNRAS.187..237B}
}

@ARTICLE{Katz:1977,
       author = {{Katz}, J.~I.},
        title = "{X-rays from spherical accretion onto degenerate dwarfs.}",
      journal = {\apj},
         year = 1977,
        month = jul,
       volume = {215},
        pages = {265-275},
          doi = {10.1086/155355},
       adsurl = {https://ui.adsabs.harvard.edu/abs/1977ApJ...215..265K}
}

@ARTICLE{Abramowicz:1988,
       author = {{Abramowicz}, M.~A. and {Czerny}, B. and {Lasota}, J.~P. and {Szuszkiewicz}, E.},
        title = "{Slim Accretion Disks}",
      journal = {\apj},
         year = 1988,
        month = sep,
       volume = {332},
        pages = {646},
          doi = {10.1086/166683},
       adsurl = {https://ui.adsabs.harvard.edu/abs/1988ApJ...332..646A}
}

@ARTICLE{RudPac:1981,
       author = {{Rudak}, B. and {Paczynski}, B.},
        title = "{Outer excretion disk around a close binary}",
      journal = {\actaa},
         year = 1981,
        month = jan,
       volume = {31},
       number = {1},
        pages = {13-24},
       adsurl = {https://ui.adsabs.harvard.edu/abs/1981AcA....31...13R}
}

@ARTICLE{Koudmani_unifiedAccModels:2024,
       author = {{Koudmani}, Sophie and {Somerville}, Rachel S. and {Sijacki}, Debora and {Bourne}, Martin A. and {Jiang}, Yan-Fei and {Profit}, Kasar},
        title = "{A unified accretion disc model for supermassive black holes in galaxy formation simulations: method and implementation}",
      journal = {\mnras},
         year = 2024,
        month = jul,
       volume = {532},
       number = {1},
        pages = {60-88},
          doi = {10.1093/mnras/stae1422},
archivePrefix = {arXiv},
       eprint = {2312.08428},
 primaryClass = {astro-ph.GA},
       adsurl = {https://ui.adsabs.harvard.edu/abs/2024MNRAS.532...60K}
}

@ARTICLE{Nemmen+2014,
       author = {{Nemmen}, Rodrigo S. and {Storchi-Bergmann}, Thaisa and {Eracleous}, Michael},
        title = "{Spectral models for low-luminosity active galactic nuclei in LINERs: the role of advection-dominated accretion and jets}",
      journal = {\mnras},
         year = 2014,
        month = mar,
       volume = {438},
       number = {4},
        pages = {2804-2827},
          doi = {10.1093/mnras/stt2388},
archivePrefix = {arXiv},
       eprint = {1312.1982},
 primaryClass = {astro-ph.HE},
       adsurl = {https://ui.adsabs.harvard.edu/abs/2014MNRAS.438.2804N}
}

@ARTICLE{Nemmen+2006,
       author = {{Nemmen}, Rodrigo S. and {Storchi-Bergmann}, Thaisa and {Yuan}, Feng and {Eracleous}, Michael and {Terashima}, Yuichi and {Wilson}, Andrew S.},
        title = "{Radiatively Inefficient Accretion Flow in the Nucleus of NGC 1097}",
      journal = {\apj},
         year = 2006,
        month = jun,
       volume = {643},
       number = {2},
        pages = {652-659},
          doi = {10.1086/500571},
archivePrefix = {arXiv},
       eprint = {astro-ph/0512540},
 primaryClass = {astro-ph},
       adsurl = {https://ui.adsabs.harvard.edu/abs/2006ApJ...643..652N}
}

@ARTICLE{SLE:1976,
       author = {{Shapiro}, S.~L. and {Lightman}, A.~P. and {Eardley}, D.~M.},
        title = "{A two-temperature accretion disk model for Cygnus X-1: structure and spectrum.}",
      journal = {\apj},
         year = 1976,
        month = feb,
       volume = {204},
        pages = {187-199},
          doi = {10.1086/154162},
       adsurl = {https://ui.adsabs.harvard.edu/abs/1976ApJ...204..187S}
}

@ARTICLE{Narayan:1996,
       author = {{Narayan}, Ramesh},
        title = "{Advection-dominated Models of Luminous Accreting Black Holes}",
      journal = {\apj},
         year = 1996,
        month = may,
       volume = {462},
        pages = {136},
          doi = {10.1086/177136},
archivePrefix = {arXiv},
       eprint = {astro-ph/9510028},
 primaryClass = {astro-ph},
       adsurl = {https://ui.adsabs.harvard.edu/abs/1996ApJ...462..136N}
}

@ARTICLE{CellaTaylorKelley:2025,
       author = {{Cella}, Katharine and {Taylor}, Stephen R. and {Zoltan Kelley}, Luke},
        title = "{Host galaxy demographics of individually detectable supermassive black-hole binaries with pulsar timing arrays}",
      journal = {Classical and Quantum Gravity},
         year = 2025,
        month = jan,
       volume = {42},
       number = {2},
          eid = {025021},
        pages = {025021},
          doi = {10.1088/1361-6382/ad9131},
archivePrefix = {arXiv},
       eprint = {2407.01659},
 primaryClass = {astro-ph.GA},
       adsurl = {https://ui.adsabs.harvard.edu/abs/2025CQGra..42b5021C}
}

@ARTICLE{Becsy_indivPTA+2025,
       author = {{B{\'e}csy}, Bence and {Cornish}, Neil J. and {Petrov}, Polina and {Siemens}, Xavier and {Taylor}, Stephen R. and {Vigeland}, Sarah J. and {Witt}, Caitlin A.},
        title = "{Towards robust gravitational wave detections from individual supermassive black hole binaries}",
      journal = {arXiv e-prints},
         year = 2025,
        month = feb,
          eid = {arXiv:2502.18114},
        pages = {arXiv:2502.18114},
          doi = {10.48550/arXiv.2502.18114},
archivePrefix = {arXiv},
       eprint = {2502.18114},
 primaryClass = {gr-qc},
       adsurl = {https://ui.adsabs.harvard.edu/abs/2025arXiv250218114B}
}

@ARTICLE{Comerford_AGNFuel+2024,
       author = {{Comerford}, Julia M. and {Nevin}, Rebecca and {Negus}, James and {Barrows}, R. Scott and {Eracleous}, Michael and {M{\"u}ller-S{\'a}nchez}, Francisco and {Roy}, Namrata and {Stemo}, Aaron and {Storchi-Bergmann}, Thaisa and {Wylezalek}, Dominika},
        title = "{An Excess of Active Galactic Nuclei Triggered by Galaxy Mergers in MaNGA Galaxies of Stellar Mass {\ensuremath{\sim}}{}10$^{11}$ M $_{{\ensuremath{\odot}}}$}",
      journal = {\apj},
         year = 2024,
        month = mar,
       volume = {963},
       number = {1},
          eid = {53},
        pages = {53},
          doi = {10.3847/1538-4357/ad1a15},
archivePrefix = {arXiv},
       eprint = {2404.14490},
 primaryClass = {astro-ph.GA},
       adsurl = {https://ui.adsabs.harvard.edu/abs/2024ApJ...963...53C}
}

@ARTICLE{Britzen_JetWiggles+2023,
       author = {{Britzen}, Silke and {Zaja{\v{c}}ek}, Michal and {Gopal-Krishna} and {Fendt}, Christian and {Kun}, Emma and {Jaron}, Fr{\'e}d{\'e}ric and {Sillanp{\"a}{\"a}}, Aimo and {Eckart}, Andreas},
        title = "{Precession-induced Variability in AGN Jets and OJ 287}",
      journal = {\apj},
         year = 2023,
        month = jul,
       volume = {951},
       number = {2},
          eid = {106},
        pages = {106},
          doi = {10.3847/1538-4357/accbbc},
archivePrefix = {arXiv},
       eprint = {2307.05838},
 primaryClass = {astro-ph.HE},
       adsurl = {https://ui.adsabs.harvard.edu/abs/2023ApJ...951..106B}
}

@ARTICLE{Roos:1988,
       author = {{Roos}, Nico},
        title = "{Jet precession in active galaxies.}",
      journal = {\apj},
         year = 1988,
        month = nov,
       volume = {334},
        pages = {95-103},
          doi = {10.1086/166820},
       adsurl = {https://ui.adsabs.harvard.edu/abs/1988ApJ...334...95R}
}

@ARTICLE{Romero+2000,
       author = {{Romero}, G.~E. and {Chajet}, L. and {Abraham}, Z. and {Fan}, J.~H.},
        title = "{Beaming and precession in the inner jet of 3C 273 --- II. The central engine}",
      journal = {\aap},
         year = 2000,
        month = aug,
       volume = {360},
        pages = {57-64},
       adsurl = {https://ui.adsabs.harvard.edu/abs/2000A&A...360...57R}
}

@ARTICLE{MohammedBog+2025,
       author = {{Mohammed}, Niana N. and {Runnoe}, Jessie C. and {Eracleous}, Michael and {Bogdanovi{\'c}}, Tamara and {Stern}, Daniel and {Simon}, Joseph and {Charisi}, Maria and {Lazio}, T. Joseph W. and {Szekerczes}, Kaitlyn and {Sigurdsson}, Steinn and {Dabbieri}, Collin},
        title = "{Testing the Hypothesis that the Quasar J0950+5128 Harbors a Supermassive Black Hole Binary}",
      journal = {arXiv e-prints},
         year = 2025,
        month = may,
          eid = {arXiv:2505.06221},
        pages = {arXiv:2505.06221},
          doi = {10.48550/arXiv.2505.06221},
archivePrefix = {arXiv},
       eprint = {2505.06221},
 primaryClass = {astro-ph.GA},
       adsurl = {https://ui.adsabs.harvard.edu/abs/2025arXiv250506221M}
}

@ARTICLE{Runnoe+2025,
       author = {{Runnoe}, Jessie C. and {Eracleous}, Michael and {Bogdanovi{\'c}}, Tamara and {Halpern}, Jules P. and {Sigur{\dj}sson}, Steinn},
        title = "{A Large Systematic Search for Close Supermassive Binary and Rapidly Recoiling Black Holes. IV. Ultraviolet Spectroscopy}",
      journal = {\apj},
         year = 2025,
        month = may,
       volume = {984},
       number = {1},
          eid = {17},
        pages = {17},
          doi = {10.3847/1538-4357/adba58},
archivePrefix = {arXiv},
       eprint = {2501.10574},
 primaryClass = {astro-ph.GA},
       adsurl = {https://ui.adsabs.harvard.edu/abs/2025ApJ...984...17R}
}

@ARTICLE{Runnoe+2017,
       author = {{Runnoe}, Jessie C. and {Eracleous}, Michael and {Pennell}, Alison and {Mathes}, Gavin and {Boroson}, Todd and {Sigur{\dh}sson}, Steinn and {Bogdanovi{\'c}}, Tamara and {Halpern}, Jules P. and {Liu}, Jia and {Brown}, Stephanie},
        title = "{A large systematic search for close supermassive binary and rapidly recoiling black holes - III. Radial velocity variations}",
      journal = {\mnras},
         year = 2017,
        month = jun,
       volume = {468},
       number = {2},
        pages = {1683-1702},
          doi = {10.1093/mnras/stx452},
archivePrefix = {arXiv},
       eprint = {1702.05465},
 primaryClass = {astro-ph.GA},
       adsurl = {https://ui.adsabs.harvard.edu/abs/2017MNRAS.468.1683R}
}

@ARTICLE{Runnoe+2015,
       author = {{Runnoe}, Jessie C. and {Eracleous}, Michael and {Mathes}, Gavin and {Pennell}, Alison and {Boroson}, Todd and {Sigur{\dh}sson}, Steinn and {Bogdanovi{\'c}}, Tamara and {Halpern}, Jules P. and {Liu}, Jia},
        title = "{A Large Systematic Search for Close Supermassive Binary and Rapidly Recoiling Black Holes. II. Continued Spectroscopic Monitoring and Optical Flux Variability}",
      journal = {\apjs},
         year = 2015,
        month = nov,
       volume = {221},
       number = {1},
          eid = {7},
        pages = {7},
          doi = {10.1088/0067-0049/221/1/710.48550/arXiv.1509.02575},
archivePrefix = {arXiv},
       eprint = {1509.02575},
 primaryClass = {astro-ph.GA},
       adsurl = {https://ui.adsabs.harvard.edu/abs/2015ApJS..221....7R}
}

@ARTICLE{Ju+2013,
       author = {{Ju}, Wenhua and {Greene}, Jenny E. and {Rafikov}, Roman R. and {Bickerton}, Steven J. and {Badenes}, Carles},
        title = "{Search for Supermassive Black Hole Binaries in the Sloan Digital Sky Survey Spectroscopic Sample}",
      journal = {\apj},
         year = 2013,
        month = nov,
       volume = {777},
       number = {1},
          eid = {44},
        pages = {44},
          doi = {10.1088/0004-637X/777/1/4410.48550/arXiv.1306.4987},
archivePrefix = {arXiv},
       eprint = {1306.4987},
 primaryClass = {astro-ph.CO},
       adsurl = {https://ui.adsabs.harvard.edu/abs/2013ApJ...777...44J}
}

@ARTICLE{Kelley_BLR:2021,
       author = {{Kelley}, Luke Zoltan},
        title = "{Basic considerations for the observability of kinematically offset binary AGN}",
      journal = {\mnras},
         year = 2021,
        month = jan,
       volume = {500},
       number = {3},
        pages = {4065-4077},
          doi = {10.1093/mnras/staa3219},
archivePrefix = {arXiv},
       eprint = {2005.10255},
 primaryClass = {astro-ph.HE},
       adsurl = {https://ui.adsabs.harvard.edu/abs/2021MNRAS.500.4065K}
}

@ARTICLE{Bogdanovic+2008,
       author = {{Bogdanovi{\'c}}, Tamara and {Smith}, Britton D. and {Sigurdsson}, Steinn and {Eracleous}, Michael},
        title = "{Modeling of Emission Signatures of Massive Black Hole Binaries. I. Methods}",
      journal = {\apjs},
         year = 2008,
        month = feb,
       volume = {174},
       number = {2},
        pages = {455-480},
          doi = {10.1086/52182810.48550/arXiv.0708.0414},
archivePrefix = {arXiv},
       eprint = {0708.0414},
 primaryClass = {astro-ph},
       adsurl = {https://ui.adsabs.harvard.edu/abs/2008ApJS..174..455B}
}

@ARTICLE{NguyenBogI:2016,
       author = {{Nguyen}, Khai and {Bogdanovi{\'c}}, Tamara},
        title = "{Emission Signatures from Sub-parsec Binary Supermassive Black Holes. I. Diagnostic Power of Broad Emission Lines}",
      journal = {\apj},
         year = 2016,
        month = sep,
       volume = {828},
       number = {2},
          eid = {68},
        pages = {68},
          doi = {10.3847/0004-637X/828/2/6810.48550/arXiv.1605.09389},
archivePrefix = {arXiv},
       eprint = {1605.09389},
 primaryClass = {astro-ph.HE},
       adsurl = {https://ui.adsabs.harvard.edu/abs/2016ApJ...828...68N}
}

@ARTICLE{NguyenBogII:2019,
       author = {{Nguyen}, Khai and {Bogdanovi{\'c}}, Tamara and {Runnoe}, Jessie C. and {Eracleous}, Michael and {Sigurdsson}, Steinn and {Boroson}, Todd},
        title = "{Emission Signatures from Sub-parsec Binary Supermassive Black Holes. II. Effect of Accretion Disk Wind on Broad Emission Lines}",
      journal = {\apj},
         year = 2019,
        month = jan,
       volume = {870},
       number = {1},
          eid = {16},
        pages = {16},
          doi = {10.3847/1538-4357/aaeff0},
archivePrefix = {arXiv},
       eprint = {1807.09782},
 primaryClass = {astro-ph.HE},
       adsurl = {https://ui.adsabs.harvard.edu/abs/2019ApJ...870...16N}
}

@ARTICLE{NguyenBogIII:2020,
       author = {{Nguyen}, Khai and {Bogdanovi{\'c}}, Tamara and {Runnoe}, Jessie C. and {Eracleous}, Michael and {Sigurdsson}, Steinn and {Boroson}, Todd},
        title = "{Emission Signatures from Subparsec Binary Supermassive Black Holes. III. Comparison of Models with Observations}",
      journal = {\apj},
         year = 2020,
        month = may,
       volume = {894},
       number = {2},
          eid = {105},
        pages = {105},
          doi = {10.3847/1538-4357/ab88b5},
archivePrefix = {arXiv},
       eprint = {1908.01799},
 primaryClass = {astro-ph.HE},
       adsurl = {https://ui.adsabs.harvard.edu/abs/2020ApJ...894..105N}
}

@ARTICLE{Ji_CBD_BLR+2021,
       author = {{Ji}, Xiang and {Lu}, Youjun and {Ge}, Junqiang and {Yan}, Changshuo and {Song}, Zihao},
        title = "{Variation of Broad Emission Lines from QSOs with Optical/UV Periodicity to Test the Interpretation of Supermassive Binary Black Holes}",
      journal = {\apj},
         year = 2021,
        month = apr,
       volume = {910},
       number = {2},
          eid = {101},
        pages = {101},
          doi = {10.3847/1538-4357/abe386},
archivePrefix = {arXiv},
       eprint = {2103.16448},
 primaryClass = {astro-ph.GA},
       adsurl = {https://ui.adsabs.harvard.edu/abs/2021ApJ...910..101J}
}

@ARTICLE{DottiPol+2022,
       author = {{Dotti}, Massimo and {Bonetti}, Matteo and {D'Orazio}, Daniel J. and {Haiman}, Zolt{\'a}n and {Ho}, Luis C.},
        title = "{Binary black hole signatures in polarized light curves}",
      journal = {\mnras},
         year = 2022,
        month = jan,
       volume = {509},
       number = {1},
        pages = {212-223},
          doi = {10.1093/mnras/stab2893},
archivePrefix = {arXiv},
       eprint = {2103.14652},
 primaryClass = {astro-ph.HE},
       adsurl = {https://ui.adsabs.harvard.edu/abs/2022MNRAS.509..212D}
}

@ARTICLE{Bertassi+2025,
       author = {{Bertassi}, Lorenzo and {Sottocorno}, Erika and {Rigamonti}, Fabio and {D'Orazio}, Daniel J. and {Eracleous}, Michael and {Haiman}, Zolt{\'a}n and {Dotti}, Massimo},
        title = "{Testing compact massive black hole binary candidates through multi-epoch spectroscopy}",
      journal = {arXiv e-prints},
         year = 2025,
        month = apr,
          eid = {arXiv:2504.06349},
        pages = {arXiv:2504.06349},
          doi = {10.48550/arXiv.2504.06349},
archivePrefix = {arXiv},
       eprint = {2504.06349},
 primaryClass = {astro-ph.GA},
       adsurl = {https://ui.adsabs.harvard.edu/abs/2025arXiv250406349B}
}

@ARTICLE{DHLighthouse+2017,
       author = {{D'Orazio}, Daniel J. and {Haiman}, Zolt{\'a}n},
        title = "{Lighthouse in the dust: infrared echoes of periodic emission from massive black hole binaries★}",
      journal = {\mnras},
         year = 2017,
        month = sep,
       volume = {470},
       number = {1},
        pages = {1198-1217},
          doi = {10.1093/mnras/stx1269},
archivePrefix = {arXiv},
       eprint = {1702.01219},
 primaryClass = {astro-ph.GA},
       adsurl = {https://ui.adsabs.harvard.edu/abs/2017MNRAS.470.1198D}
}

@ARTICLE{Caliskan_GaiaGW+2024,
       author = {{{\c{C}}al{\i}{\c{s}}kan}, Mesut and {Chen}, Yifan and {Dai}, Liang and {Kumar}, Neha Anil and {Stomberg}, Isak and {Xue}, Xiao},
        title = "{Dissecting the stochastic gravitational wave background with astrometry}",
      journal = {\jcap},
         year = 2024,
        month = may,
       volume = {2024},
       number = {5},
          eid = {030},
        pages = {030},
          doi = {10.1088/1475-7516/2024/05/030},
archivePrefix = {arXiv},
       eprint = {2312.03069},
 primaryClass = {gr-qc},
       adsurl = {https://ui.adsabs.harvard.edu/abs/2024JCAP...05..030C}
}

@ARTICLE{Foster_muHz_binres+2025,
       author = {{Foster}, Joshua W. and {Blas}, Diego and {Bourgoin}, Adrien and {Hees}, Aurelien and {Herrero-Valea}, M{\'\i}riam and {Jenkins}, Alexander C. and {Xue}, Xiao},
        title = "{Discovering $μ$Hz gravitational waves and ultra-light dark matter with binary resonances}",
      journal = {arXiv e-prints},
         year = 2025,
        month = apr,
          eid = {arXiv:2504.15334},
        pages = {arXiv:2504.15334},
          doi = {10.48550/arXiv.2504.15334},
archivePrefix = {arXiv},
       eprint = {2504.15334},
 primaryClass = {astro-ph.CO},
       adsurl = {https://ui.adsabs.harvard.edu/abs/2025arXiv250415334F}
}

@ARTICLE{ZwickUOP+2025,
       author = {{Zwick}, Lorenz and {Soyuer}, Deniz and {D'Orazio}, Daniel J. and {O'Neill}, David and {Derdzinski}, Andrea and {Saha}, Prasenjit and {Blas}, Diego and {Jenkins}, Alexander C. and {Kelley}, Luke Zoltan},
        title = "{Bridging the micro-Hz gravitational wave gap via Doppler tracking with the Uranus Orbiter and Probe Mission: Massive black hole binaries, early universe signals and ultra-light dark matter}",
      journal = {arXiv e-prints},
         year = 2024,
        month = jun,
          eid = {arXiv:2406.02306},
        pages = {arXiv:2406.02306},
          doi = {10.48550/arXiv.2406.02306},
archivePrefix = {arXiv},
       eprint = {2406.02306},
 primaryClass = {astro-ph.HE},
       adsurl = {https://ui.adsabs.harvard.edu/abs/2024arXiv240602306Z}
}

@ARTICLE{Esin+1997,
       author = {{Esin}, Ann A. and {McClintock}, Jeffrey E. and {Narayan}, Ramesh},
        title = "{Advection-Dominated Accretion and the Spectral States of Black Hole X-Ray Binaries: Application to Nova Muscae 1991}",
      journal = {\apj},
         year = 1997,
        month = nov,
       volume = {489},
       number = {2},
        pages = {865-889},
          doi = {10.1086/304829},
archivePrefix = {arXiv},
       eprint = {astro-ph/9705237},
 primaryClass = {astro-ph},
       adsurl = {https://ui.adsabs.harvard.edu/abs/1997ApJ...489..865E}
}

@ARTICLE{Menou_rtr+1999,
       author = {{Menou}, Kristen and {Narayan}, Ramesh and {Lasota}, Jean-Pierre},
        title = "{A Population of Faint Nontransient Low-Mass Black Hole Binaries}",
      journal = {\apj},
         year = 1999,
        month = mar,
       volume = {513},
       number = {2},
        pages = {811-826},
          doi = {10.1086/306878},
archivePrefix = {arXiv},
       eprint = {astro-ph/9810186},
 primaryClass = {astro-ph},
       adsurl = {https://ui.adsabs.harvard.edu/abs/1999ApJ...513..811M}
}

@ARTICLE{LISA_LRR_2023,
       author = {{Amaro-Seoane}, Pau and {Andrews}, Jeff and {Arca Sedda}, Manuel and {Askar}, Abbas and {Baghi}, Quentin and {Balasov}, Razvan and {Bartos}, Imre and {Bavera}, Simone S. and {Bellovary}, Jillian and {Berry}, Christopher P.~L. and {Berti}, Emanuele and {Bianchi}, Stefano and {Blecha}, Laura and {Blondin}, St{\'e}phane and {Bogdanovi{\'c}}, Tamara and {Boissier}, Samuel and {Bonetti}, Matteo and {Bonoli}, Silvia and {Bortolas}, Elisa and {Breivik}, Katelyn and {Capelo}, Pedro R. and {Caramete}, Laurentiu and {Cattorini}, Federico and {Charisi}, Maria and {Chaty}, Sylvain and {Chen}, Xian and {Chru{\'s}li{\'n}ska}, Martyna and {Chua}, Alvin J.~K. and {Church}, Ross and {Colpi}, Monica and {D'Orazio}, Daniel and {Danielski}, Camilla and {Davies}, Melvyn B. and {Dayal}, Pratika and {De Rosa}, Alessandra and {Derdzinski}, Andrea and {Destounis}, Kyriakos and {Dotti}, Massimo and {Du{\c{t}}an}, Ioana and {Dvorkin}, Irina and {Fabj}, Gaia and {Foglizzo}, Thierry and {Ford}, Saavik and {Fouvry}, Jean-Baptiste and {Franchini}, Alessia and {Fragos}, Tassos and {Fryer}, Chris and {Gaspari}, Massimo and {Gerosa}, Davide and {Graziani}, Luca and {Groot}, Paul and {Habouzit}, Melanie and {Haggard}, Daryl and {Haiman}, Zoltan and {Han}, Wen-Biao and {Istrate}, Alina and {Johansson}, Peter H. and {Khan}, Fazeel Mahmood and {Kimpson}, Tomas and {Kokkotas}, Kostas and {Kong}, Albert and {Korol}, Valeriya and {Kremer}, Kyle and {Kupfer}, Thomas and {Lamberts}, Astrid and {Larson}, Shane and {Lau}, Mike and {Liu}, Dongliang and {Lloyd-Ronning}, Nicole and {Lodato}, Giuseppe and {Lupi}, Alessandro and {Ma}, Chung-Pei and {Maccarone}, Tomas and {Mandel}, Ilya and {Mangiagli}, Alberto and {Mapelli}, Michela and {Mathis}, St{\'e}phane and {Mayer}, Lucio and {McGee}, Sean and {McKernan}, Berry and {Miller}, M. Coleman and {Mota}, David F. and {Mumpower}, Matthew and {Nasim}, Syeda S. and {Nelemans}, Gijs and {Noble}, Scott and {Pacucci}, Fabio and {Panessa}, Francesca and {Paschalidis}, Vasileios and {Pfister}, Hugo and {Porquet}, Delphine and {Quenby}, John and {Ricarte}, Angelo and {R{\"o}pke}, Friedrich K. and {Regan}, John and {Rosswog}, Stephan and {Ruiter}, Ashley and {Ruiz}, Milton and {Runnoe}, Jessie and {Schneider}, Raffaella and {Schnittman}, Jeremy and {Secunda}, Amy and {Sesana}, Alberto and {Seto}, Naoki and {Shao}, Lijing and {Shapiro}, Stuart and {Sopuerta}, Carlos and {Stone}, Nicholas C. and {Suvorov}, Arthur and {Tamanini}, Nicola and {Tamfal}, Tomas and {Tauris}, Thomas and {Temmink}, Karel and {Tomsick}, John and {Toonen}, Silvia and {Torres-Orjuela}, Alejandro and {Toscani}, Martina and {Tsokaros}, Antonios and {Unal}, Caner and {V{\'a}zquez-Aceves}, Ver{\'o}nica and {Valiante}, Rosa and {van Putten}, Maurice and {van Roestel}, Jan and {Vignali}, Christian and {Volonteri}, Marta and {Wu}, Kinwah and {Younsi}, Ziri and {Yu}, Shenghua and {Zane}, Silvia and {Zwick}, Lorenz and {Antonini}, Fabio and {Baibhav}, Vishal and {Barausse}, Enrico and {Bonilla Rivera}, Alexander and {Branchesi}, Marica and {Branduardi-Raymont}, Graziella and {Burdge}, Kevin and {Chakraborty}, Srija and {Cuadra}, Jorge and {Dage}, Kristen and {Davis}, Benjamin and {de Mink}, Selma E. and {Decarli}, Roberto and {Doneva}, Daniela and {Escoffier}, Stephanie and {Gandhi}, Poshak and {Haardt}, Francesco and {Lousto}, Carlos O. and {Nissanke}, Samaya and {Nordhaus}, Jason and {O'Shaughnessy}, Richard and {Portegies Zwart}, Simon and {Pound}, Adam and {Schussler}, Fabian and {Sergijenko}, Olga and {Spallicci}, Alessandro and {Vernieri}, Daniele and {Vigna-G{\'o}mez}, Alejandro},
        title = "{Astrophysics with the Laser Interferometer Space Antenna}",
      journal = {Living Reviews in Relativity},
         year = 2023,
        month = dec,
       volume = {26},
       number = {1},
          eid = {2},
        pages = {2},
          doi = {10.1007/s41114-022-00041-y},
archivePrefix = {arXiv},
       eprint = {2203.06016},
 primaryClass = {gr-qc},
       adsurl = {https://ui.adsabs.harvard.edu/abs/2023LRR....26....2A}
}

@ARTICLE{Gardiner+25,
       author = {{Gardiner}, Emiko C. and {B{\'e}csy}, Bence and {Kelley}, Luke Zoltan and {Cornish}, Neil J.},
        title = "{Characterizing Continuous Gravitational Waves from Supermassive Black Hole Binaries in Realistic Pulsar Timing Array Data}",
      journal = {\apj},
         year = 2025,
        month = aug,
       volume = {988},
       number = {2},
          eid = {222},
        pages = {222},
          doi = {10.3847/1538-4357/ade4c2},
archivePrefix = {arXiv},
       eprint = {2502.16016},
 primaryClass = {astro-ph.CO},
       adsurl = {https://ui.adsabs.harvard.edu/abs/2025ApJ...988..222G}
}

@ARTICLE{Miles+25,
       author = {{Miles}, Matthew T. and {Shannon}, Ryan M. and {Reardon}, Daniel J. and {Bailes}, Matthew and {Champion}, David J. and {Geyer}, Marisa and {Gitika}, Pratyasha and {Grunthal}, Kathrin and {Keith}, Michael J. and {Kramer}, Michael and {Kulkarni}, Atharva D. and {Nathan}, Rowina S. and {Parthasarathy}, Aditya and {Singha}, Jaikhomba and {Theureau}, Gilles and {Thrane}, Eric and {Abbate}, Federico and {Buchner}, Sarah and {Cameron}, Andrew D. and {Camilo}, Fernando and {Moreschi}, Beatrice E. and {Shaifullah}, Golam and {Shamohammadi}, Mohsen and {Possenti}, Andrea and {Krishnan}, Vivek Venkatraman},
        title = "{The MeerKAT Pulsar Timing Array: the first search for gravitational waves with the MeerKAT radio telescope}",
      journal = {\mnras},
         year = 2025,
        month = jan,
       volume = {536},
       number = {2},
        pages = {1489-1500},
          doi = {10.1093/mnras/stae2571},
archivePrefix = {arXiv},
       eprint = {2412.01153},
 primaryClass = {astro-ph.HE},
       adsurl = {https://ui.adsabs.harvard.edu/abs/2025MNRAS.536.1489M}
}

@ARTICLE{iPTA2022,
       author = {{Antoniadis}, J. and {Arzoumanian}, Z. and {Babak}, S. and {Bailes}, M. and {Bak Nielsen}, A. -S. and {Baker}, P.~T. and {Bassa}, C.~G. and {B{\'e}csy}, B. and {Berthereau}, A. and {Bonetti}, M. and {Brazier}, A. and {Brook}, P.~R. and {Burgay}, M. and {Burke-Spolaor}, S. and {Caballero}, R.~N. and {Casey-Clyde}, J.~A. and {Chalumeau}, A. and {Champion}, D.~J. and {Charisi}, M. and {Chatterjee}, S. and {Chen}, S. and {Cognard}, I. and {Cordes}, J.~M. and {Cornish}, N.~J. and {Crawford}, F. and {Cromartie}, H.~T. and {Crowter}, K. and {Dai}, S. and {DeCesar}, M.~E. and {Demorest}, P.~B. and {Desvignes}, G. and {Dolch}, T. and {Drachler}, B. and {Falxa}, M. and {Ferrara}, E.~C. and {Fiore}, W. and {Fonseca}, E. and {Gair}, J.~R. and {Garver-Daniels}, N. and {Goncharov}, B. and {Good}, D.~C. and {Graikou}, E. and {Guillemot}, L. and {Guo}, Y.~J. and {Hazboun}, J.~S. and {Hobbs}, G. and {Hu}, H. and {Islo}, K. and {Janssen}, G.~H. and {Jennings}, R.~J. and {Johnson}, A.~D. and {Jones}, M.~L. and {Kaiser}, A.~R. and {Kaplan}, D.~L. and {Karuppusamy}, R. and {Keith}, M.~J. and {Kelley}, L.~Z. and {Kerr}, M. and {Key}, J.~S. and {Kramer}, M. and {Lam}, M.~T. and {Lamb}, W.~G. and {Lazio}, T.~J.~W. and {Lee}, K.~J. and {Lentati}, L. and {Liu}, K. and {Luo}, J. and {Lynch}, R.~S. and {Lyne}, A.~G. and {Madison}, D.~R. and {Main}, R.~A. and {Manchester}, R.~N. and {McEwen}, A. and {McKee}, J.~W. and {McLaughlin}, M.~A. and {Mickaliger}, M.~B. and {Mingarelli}, C.~M.~F. and {Ng}, C. and {Nice}, D.~J. and {Os{\l}owski}, S. and {Parthasarathy}, A. and {Pennucci}, T.~T. and {Perera}, B.~B.~P. and {Perrodin}, D. and {Petiteau}, A. and {Pol}, N.~S. and {Porayko}, N.~K. and {Possenti}, A. and {Ransom}, S.~M. and {Ray}, P.~S. and {Reardon}, D.~J. and {Russell}, C.~J. and {Samajdar}, A. and {Sampson}, L.~M. and {Sanidas}, S. and {Sarkissian}, J.~M. and {Schmitz}, K. and {Schult}, L. and {Sesana}, A. and {Shaifullah}, G. and {Shannon}, R.~M. and {Shapiro-Albert}, B.~J. and {Siemens}, X. and {Simon}, J. and {Smith}, T.~L. and {Speri}, L. and {Spiewak}, R. and {Stairs}, I.~H. and {Stappers}, B.~W. and {Stinebring}, D.~R. and {Swiggum}, J.~K. and {Taylor}, S.~R. and {Theureau}, G. and {Tiburzi}, C. and {Vallisneri}, M. and {van der Wateren}, E. and {Vecchio}, A. and {Verbiest}, J.~P.~W. and {Vigeland}, S.~J. and {Wahl}, H. and {Wang}, J.~B. and {Wang}, J. and {Wang}, L. and {Witt}, C.~A. and {Zhang}, S. and {Zhu}, X.~J.},
        title = "{The International Pulsar Timing Array second data release: Search for an isotropic gravitational wave background}",
      journal = {\mnras},
         year = 2022,
        month = mar,
       volume = {510},
       number = {4},
        pages = {4873-4887},
          doi = {10.1093/mnras/stab3418},
archivePrefix = {arXiv},
       eprint = {2201.03980},
 primaryClass = {astro-ph.HE},
       adsurl = {https://ui.adsabs.harvard.edu/abs/2022MNRAS.510.4873A}
}

@ARTICLE{SesanaVecchio+2008,
       author = {{Sesana}, A. and {Vecchio}, A. and {Colacino}, C.~N.},
        title = "{The stochastic gravitational-wave background from massive black hole binary systems: implications for observations with Pulsar Timing Arrays}",
      journal = {\mnras},
         year = 2008,
        month = oct,
       volume = {390},
       number = {1},
        pages = {192-209},
          doi = {10.1111/j.1365-2966.2008.13682.x},
archivePrefix = {arXiv},
       eprint = {0804.4476},
 primaryClass = {astro-ph},
       adsurl = {https://ui.adsabs.harvard.edu/abs/2008MNRAS.390..192S}
}

@ARTICLE{Carr:1980,
       author = {{Carr}, B.~J.},
        title = "{Cosmological gravitational waves - Their origin and consequences}",
      journal = {\aap},
         year = 1980,
        month = sep,
       volume = {89},
       number = {1-2},
        pages = {6-21},
       adsurl = {https://ui.adsabs.harvard.edu/abs/1980A&A....89....6C}
}

@ARTICLE{Franchini_eccSG+2024,
       author = {{Franchini}, Alessia and {Prato}, Alessandra and {Longarini}, Cristiano and {Sesana}, Alberto},
        title = "{The behaviour of eccentric sub-pc massive black hole binaries embedded in massive discs}",
      journal = {\aap},
         year = 2024,
        month = aug,
       volume = {688},
          eid = {A174},
        pages = {A174},
          doi = {10.1051/0004-6361/202449402},
archivePrefix = {arXiv},
       eprint = {2402.00938},
 primaryClass = {astro-ph.HE},
       adsurl = {https://ui.adsabs.harvard.edu/abs/2024A&A...688A.174F}
}

@ARTICLE{ZhuThrane:2020,
       author = {{Zhu}, Xing-Jiang and {Thrane}, Eric},
        title = "{Toward the Unambiguous Identification of Supermassive Binary Black Holes through Bayesian Inference}",
      journal = {\apj},
         year = 2020,
        month = sep,
       volume = {900},
       number = {2},
          eid = {117},
        pages = {117},
          doi = {10.3847/1538-4357/abac5a},
archivePrefix = {arXiv},
       eprint = {2004.10944},
 primaryClass = {astro-ph.HE},
       adsurl = {https://ui.adsabs.harvard.edu/abs/2020ApJ...900..117Z}
}

@ARTICLE{Charisi+2018,
       author = {{Charisi}, Maria and {Haiman}, Zolt{\'a}n and {Schiminovich}, David and {D'Orazio}, Daniel J.},
        title = "{Testing the relativistic Doppler boost hypothesis for supermassive black hole binary candidates}",
      journal = {\mnras},
         year = 2018,
        month = jun,
       volume = {476},
       number = {4},
        pages = {4617-4628},
          doi = {10.1093/mnras/sty516},
archivePrefix = {arXiv},
       eprint = {1801.06189},
 primaryClass = {astro-ph.GA},
       adsurl = {https://ui.adsabs.harvard.edu/abs/2018MNRAS.476.4617C}
}

@ARTICLE{KelleyLens+2021,
       author = {{Kelley}, Luke Zoltan and {D'Orazio}, Daniel J. and {Di Stefano}, Rosanne},
        title = "{Gravitational self-lensing in populations of massive black hole binaries}",
      journal = {\mnras},
         year = 2021,
        month = dec,
       volume = {508},
       number = {2},
        pages = {2524-2536},
          doi = {10.1093/mnras/stab2776},
archivePrefix = {arXiv},
       eprint = {2107.07522},
 primaryClass = {astro-ph.HE},
       adsurl = {https://ui.adsabs.harvard.edu/abs/2021MNRAS.508.2524K}
}

@ARTICLE{XinHaiman_LSSTshort:2021,
       author = {{Xin}, Chengcheng and {Haiman}, Zolt{\'a}n},
        title = "{Ultra-short-period massive black hole binary candidates in LSST as LISA 'verification binaries'}",
      journal = {\mnras},
         year = 2021,
        month = sep,
       volume = {506},
       number = {2},
        pages = {2408-2417},
          doi = {10.1093/mnras/stab1856},
archivePrefix = {arXiv},
       eprint = {2105.00005},
 primaryClass = {astro-ph.HE},
       adsurl = {https://ui.adsabs.harvard.edu/abs/2021MNRAS.506.2408X}
}

@ARTICLE{RomanPLCs+2023,
       author = {{Haiman}, Zolt{\'a}n and {Xin}, Chengcheng and {Bogdanovi{\'c}}, Tamara and {Amaro Seoane}, Pau and {Bonetti}, Matteo and {Casey-Clyde}, J. Andrew and {Charisi}, Maria and {Colpi}, Monica and {Davelaar}, Jordy and {De Rosa}, Alessandra and {D'Orazio}, Daniel J. and {Futrowsky}, Kate and {Gandhi}, Poshak and {Graham}, Alister W. and {Greene}, Jenny E. and {Habouzit}, Melanie and {Haggard}, Daryl and {Holley-Bockelmann}, Kelly and {Liu}, Xin and {Mangiagli}, Alberto and {Mastrobuono-Battisti}, Alessandra and {McGee}, Sean and {Mingarelli}, Chiara M.~F. and {Nemmen}, Rodrigo and {Palmese}, Antonella and {Porquet}, Delphine and {Sesana}, Alberto and {Stemo}, Aaron and {Torres-Orjuela}, Alejandro and {Zrake}, Jonathan},
        title = "{Massive Black Hole Binaries as LISA Precursors in the Roman High Latitude Time Domain Survey}",
      journal = {arXiv e-prints},
         year = 2023,
        month = jun,
          eid = {arXiv:2306.14990},
        pages = {arXiv:2306.14990},
          doi = {10.48550/arXiv.2306.14990},
archivePrefix = {arXiv},
       eprint = {2306.14990},
 primaryClass = {astro-ph.HE},
       adsurl = {https://ui.adsabs.harvard.edu/abs/2023arXiv230614990H}
}

@ARTICLE{dAscoli+2018,
       author = {{d'Ascoli}, St{\'e}phane and {Noble}, Scott C. and {Bowen}, Dennis B. and {Campanelli}, Manuela and {Krolik}, Julian H. and {Mewes}, Vassilios},
        title = "{Electromagnetic Emission from Supermassive Binary Black Holes Approaching Merger}",
      journal = {\apj},
         year = 2018,
        month = oct,
       volume = {865},
       number = {2},
          eid = {140},
        pages = {140},
          doi = {10.3847/1538-4357/aad8b410.48550/arXiv.1806.05697},
archivePrefix = {arXiv},
       eprint = {1806.05697},
 primaryClass = {astro-ph.HE},
       adsurl = {https://ui.adsabs.harvard.edu/abs/2018ApJ...865..140D}
}

@ARTICLE{DOrazioCharisi:2023,
       author = {{D'Orazio}, Daniel J. and {Charisi}, Maria},
        title = "{Observational Signatures of Supermassive Black Hole Binaries}",
      journal = {arXiv e-prints},
         year = 2023,
        month = oct,
          eid = {arXiv:2310.16896},
        pages = {arXiv:2310.16896},
          doi = {10.48550/arXiv.2310.16896},
archivePrefix = {arXiv},
       eprint = {2310.16896},
 primaryClass = {astro-ph.HE},
       adsurl = {https://ui.adsabs.harvard.edu/abs/2023arXiv231016896D}
}

@ARTICLE{MaheshMcW+2024,
       author = {{Mahesh}, Siddharth and {McWilliams}, Sean T. and {Pirog}, Michal},
        title = "{Analytical and Numerical Analysis of Circumbinary Disk Dynamics. I. Coplanar Systems}",
      journal = {\apj},
         year = 2024,
        month = sep,
       volume = {973},
       number = {1},
          eid = {18},
        pages = {18},
          doi = {10.3847/1538-4357/ad6149},
archivePrefix = {arXiv},
       eprint = {2305.01533},
 primaryClass = {astro-ph.SR},
       adsurl = {https://ui.adsabs.harvard.edu/abs/2024ApJ...973...18M}
}

@ARTICLE{Roedig_SEDsigs+2014,
       author = {{Roedig}, Constanze and {Krolik}, Julian H. and {Miller}, M. Coleman},
        title = "{Observational Signatures of Binary Supermassive Black Holes}",
      journal = {\apj},
         year = 2014,
        month = apr,
       volume = {785},
       number = {2},
          eid = {115},
        pages = {115},
          doi = {10.1088/0004-637X/785/2/115},
archivePrefix = {arXiv},
       eprint = {1402.7098},
 primaryClass = {astro-ph.HE},
       adsurl = {https://ui.adsabs.harvard.edu/abs/2014ApJ...785..115R}
}

@ARTICLE{GultekinMiller_SEDGaps:2012,
       author = {{G{\"u}ltekin}, Kayhan and {Miller}, Jon M.},
        title = "{Observable Consequences of Merger-driven Gaps and Holes in Black Hole Accretion Disks}",
      journal = {\apj},
         year = 2012,
        month = dec,
       volume = {761},
       number = {2},
          eid = {90},
        pages = {90},
          doi = {10.1088/0004-637X/761/2/90},
archivePrefix = {arXiv},
       eprint = {1207.0296},
 primaryClass = {astro-ph.HE},
       adsurl = {https://ui.adsabs.harvard.edu/abs/2012ApJ...761...90G}
}

@ARTICLE{Zrake+2021,
       author = {{Zrake}, Jonathan and {Tiede}, Christopher and {MacFadyen}, Andrew and {Haiman}, Zolt{\'a}n},
        title = "{Equilibrium Eccentricity of Accreting Binaries}",
      journal = {\apjl},
         year = 2021,
        month = mar,
       volume = {909},
       number = {1},
          eid = {L13},
        pages = {L13},
          doi = {10.3847/2041-8213/abdd1c},
archivePrefix = {arXiv},
       eprint = {2010.09707},
 primaryClass = {astro-ph.HE},
       adsurl = {https://ui.adsabs.harvard.edu/abs/2021ApJ...909L..13Z}
}

@ARTICLE{Duffell:2020,
       author = {{Duffell}, Paul C. and {D'Orazio}, Daniel and {Derdzinski}, Andrea and {Haiman}, Zoltan and {MacFadyen}, Andrew and {Rosen}, Anna L. and {Zrake}, Jonathan},
        title = "{Circumbinary Disks: Accretion and Torque as a Function of Mass Ratio and Disk Viscosity}",
      journal = {\apj},
         year = 2020,
        month = sep,
       volume = {901},
       number = {1},
          eid = {25},
        pages = {25},
          doi = {10.3847/1538-4357/abab95},
archivePrefix = {arXiv},
       eprint = {1911.05506},
 primaryClass = {astro-ph.SR},
       adsurl = {https://ui.adsabs.harvard.edu/abs/2020ApJ...901...25D}
}

@ARTICLE{DOrazioDuffell:2021,
       author = {{D'Orazio}, Daniel J. and {Duffell}, Paul C.},
        title = "{Orbital Evolution of Equal-mass Eccentric Binaries due to a Gas Disk: Eccentric Inspirals and Circular Outspirals}",
      journal = {\apjl},
         year = 2021,
        month = jun,
       volume = {914},
       number = {1},
          eid = {L21},
        pages = {L21},
          doi = {10.3847/2041-8213/ac0621},
archivePrefix = {arXiv},
       eprint = {2103.09251},
 primaryClass = {astro-ph.HE},
       adsurl = {https://ui.adsabs.harvard.edu/abs/2021ApJ...914L..21D}
}

@ARTICLE{Ostriker:99,
       author = {{Ostriker}, Eve C.},
        title = "{Dynamical Friction in a Gaseous Medium}",
      journal = {\apj},
         year = 1999,
        month = mar,
       volume = {513},
       number = {1},
        pages = {252-258},
          doi = {10.1086/306858},
archivePrefix = {arXiv},
       eprint = {astro-ph/9810324},
 primaryClass = {astro-ph},
       adsurl = {https://ui.adsabs.harvard.edu/abs/1999ApJ...513..252O}
}

@ARTICLE{DOrazioLoebVLBI:2018,
       author = {{D'Orazio}, Daniel J. and {Loeb}, Abraham},
        title = "{Repeated Imaging of Massive Black Hole Binary Orbits with Millimeter Interferometry: Measuring Black Hole Masses and the Hubble Constant}",
      journal = {\apj},
         year = 2018,
        month = aug,
       volume = {863},
       number = {2},
          eid = {185},
        pages = {185},
          doi = {10.3847/1538-4357/aad413},
archivePrefix = {arXiv},
       eprint = {1712.02362},
 primaryClass = {astro-ph.HE},
       adsurl = {https://ui.adsabs.harvard.edu/abs/2018ApJ...863..185D}
}

@article{AL94,
author = {Pawel Artymowicz and Stephen H Lubow}, 
journal = {\apj},
title = {Dynamics of binary-disk interaction. 1: Resonances and disk gap sizes},
affiliation = {AA(University of California, Santa Cruz, CA, US), AB(University of California, Santa Cruz, CA, US)},
pages = {651},
volume = {421},
year = {1994},
month = {Feb},
doi = {10.1086/173679},
pmid = {1994ApJ...421..651A},
uri = {papers://77183D9B-B140-4DA6-B598-A9837E5CF53B/Paper/p1011}
}

@article{MacFadyen:2008,
author = {Andrew I MacFadyen and Milo{\v s} Milosavljevi{\'c}}, 
journal = {\apj},
title = {An Eccentric Circumbinary Accretion Disk and the Detection of Binary Massive Black Holes},
affiliation = {AA(Institute for Advanced Study, Einstein Drive, Princeton, NJ 08540.; Department of Physics, New York University, New York, NY 10003.), AB(Theoretical Astrophysics, California Institute of Technology, 1200 East California Boulevard, Mail Code 130-33, Pasadena, CA 91125.; Hubble Fellow.; Department of Astronomy, University of Texas, 1 University Station C1400, Austin, TX 78712.)},
pages = {83},
volume = {672},
year = {2008},
month = {Jan},
doi = {10.1086/523869},
pmid = {2008ApJ...672...83M},
uri = {papers://77183D9B-B140-4DA6-B598-A9837E5CF53B/Paper/p80}
}

@ARTICLE{Begel:Blan:Rees:1980,
   author = {{Begelman}, M.~C. and {Blandford}, R.~D. and {Rees}, M.~J.},
    title = "{Massive black hole binaries in active galactic nuclei}",
  journal = {\nat},
     year = 1980,
    month = sep,
   volume = 287,
    pages = {307-309},
      doi = {10.1038/287307a0},
}

@article{Roos:1993,
author = {Nico Roos and Jelle S Kaastra and Christian A Hummel}, 
journal = {\apj},
title = {A massive binary black hole in 1928 + 738?},
affiliation = {AA(Leiden, Sterrewacht, Netherlands), AB(SRON, Leiden, Netherlands), AC(Universities Space Research Association, Washington)},
pages = {130},
volume = {409},
year = {1993},
month = {May},
doi = {10.1086/172647},
pmid = {1993ApJ...409..130R},
uri = {papers://77183D9B-B140-4DA6-B598-A9837E5CF53B/Paper/p1792}
}

@ARTICLE{SS73,
   author = {{Shakura}, N.~I. and {Sunyaev}, R.~A.},
    title = "{Black holes in binary systems. Observational appearance.}",
  journal = {\aap},
     year = 1973,
   volume = 24,
    pages = {337-355},
   adsurl = {http://adsabs.harvard.edu/abs/1973A%26A....24..337S}
}

@ARTICLE{Haiman+2009,
   author = {{Haiman}, Z. and {Kocsis}, B. and {Menou}, K. and {Lippai}, Z. and 
	{Frei}, Z.},
    title = "{Identifying decaying supermassive black hole binaries from their variable electromagnetic emission}",
  journal = {Classical and Quantum Gravity},
archivePrefix = "arXiv",
     year = 2009,
    month = may,
   volume = 26,
   number = 9,
    pages = {094032},
      doi = {10.1088/0264-9381/26/9/094032},
}

@BOOK{Rybicki:RadiativeProcesses,
       author = {{Rybicki}, George B. and {Lightman}, Alan P.},
        title = "{Radiative Processes in Astrophysics}",
         year = 1986,
       adsurl = {https://ui.adsabs.harvard.edu/abs/1986rpa..book.....R}
}

@ARTICLE{FarrisLiuShap:2010:Bondi,
   author = {{Farris}, B.~D. and {Liu}, Y.~T. and {Shapiro}, S.~L.},
    title = "{Binary black hole mergers in gaseous environments: ``Binary Bondi'' and ``binary Bondi-Hoyle-Lyttleton'' accretion}",
  journal = {\prd},
archivePrefix = "arXiv",
 primaryClass = "astro-ph.HE",
     year = 2010,
    month = apr,
   volume = 81,
   number = 8,
      eid = {084008},
    pages = {084008},
      doi = {10.1103/PhysRevD.81.084008},
}

@ARTICLE{Bode:2010,
       author = {{Bode}, Tanja and {Haas}, Roland and {Bogdanovi{\'c}}, Tamara and {Laguna}, Pablo and {Shoemaker}, Deirdre},
        title = "{Relativistic Mergers of Supermassive Black Holes and Their Electromagnetic Signatures}",
      journal = {\apj},
         year = 2010,
        month = jun,
       volume = {715},
       number = {2},
        pages = {1117-1131},
          doi = {10.1088/0004-637X/715/2/1117},
archivePrefix = {arXiv},
       eprint = {0912.0087},
 primaryClass = {gr-qc},
       adsurl = {https://ui.adsabs.harvard.edu/abs/2010ApJ...715.1117B}
}

@ARTICLE{Bode:2012,
       author = {{Bode}, Tanja and {Bogdanovi{\'c}}, Tamara and {Haas}, Roland and {Healy}, James and {Laguna}, Pablo and {Shoemaker}, Deirdre},
        title = "{Mergers of Supermassive Black Holes in Astrophysical Environments}",
      journal = {\apj},
         year = 2012,
        month = jan,
       volume = {744},
       number = {1},
          eid = {45},
        pages = {45},
          doi = {10.1088/0004-637X/744/1/45},
archivePrefix = {arXiv},
       eprint = {1101.4684},
 primaryClass = {gr-qc},
       adsurl = {https://ui.adsabs.harvard.edu/abs/2012ApJ...744...45B}
}

@ARTICLE{DHM:2013:MNRAS,
   author = {{D'Orazio}, D.~J. and {Haiman}, Z. and {MacFadyen}, A.},
    title = "{Accretion into the central cavity of a circumbinary disc}",
  journal = {\mnras},
archivePrefix = "arXiv",
 primaryClass = "astro-ph.GA",
     year = 2013,
    month = dec,
   volume = 436,
    pages = {2997-3020},
      doi = {10.1093/mnras/stt1787},
}

@ARTICLE{Farris:2014,
   author = {{Farris}, B.~D. and {Duffell}, P. and {MacFadyen}, A.~I. and 
	{Haiman}, Z.},
    title = "{Binary Black Hole Accretion from a Circumbinary Disk: Gas Dynamics inside the Central Cavity}",
  journal = {\apj},
archivePrefix = "arXiv",
   eprint = {1310.0492},
 primaryClass = "astro-ph.HE",
     year = 2014,
    month = mar,
   volume = 783,
      eid = {134},
    pages = {134},
      doi = {10.1088/0004-637X/783/2/134},
   adsurl = {http://adsabs.harvard.edu/abs/2014ApJ...783..134F}
}

@ARTICLE{McKFeZoltan:2013,
   author = {{McKernan}, B. and {Ford}, K.~E.~S. and {Kocsis}, B. and {Haiman}, Z.
	},
    title = "{Ripple effects and oscillations in the broad Fe K{$\alpha$} line as a probe of massive black hole mergers}",
  journal = {\mnras},
archivePrefix = "arXiv",
 primaryClass = "astro-ph.HE",
     year = 2013,
    month = jun,
   volume = 432,
    pages = {1468-1482},
      doi = {10.1093/mnras/stt567},
}

@ARTICLE{LBPringle:1974,
   author = {{Lynden-Bell}, D. and {Pringle}, J.~E.},
    title = "{The evolution of viscous discs and the origin of the nebular variables.}",
  journal = {\mnras},
     year = 1974,
    month = sep,
   volume = 168,
    pages = {603-637},
}

@ARTICLE{Paczynski:1977,
   author = {{Paczynski}, B.},
    title = "{A model of accretion disks in close binaries}",
  journal = {\apj},
     year = 1977,
    month = sep,
   volume = 216,
    pages = {822-826},
      doi = {10.1086/155526},
   adsurl = {http://adsabs.harvard.edu/abs/1977ApJ...216..822P}
}

@ARTICLE{Farris:2015:Cool,
   author = {{Farris}, B.~D. and {Duffell}, P. and {MacFadyen}, A.~I. and 
	{Haiman}, Z.},
    title = "{Characteristic signatures in the thermal emission from accreting binary black holes}",
  journal = {\mnras},
archivePrefix = "arXiv",
   eprint = {1406.0007},
 primaryClass = "astro-ph.HE",
     year = 2015,
    month = jan,
   volume = 446,
    pages = {L36-L40},
      doi = {10.1093/mnrasl/slu160},
   adsurl = {http://adsabs.harvard.edu/abs/2015MNRAS.446L..36F}
}

@ARTICLE{PG1302MNRAS:2015a,
   author = {{D'Orazio}, D.~J. and {Haiman}, Z. and {Duffell}, P. and {Farris}, B.~D. and 
	{MacFadyen}, A.~I.},
    title = "{A reduced orbital period for the supermassive black hole binary candidate in the quasar PG 1302-102?}",
  journal = {\mnras},
archivePrefix = "arXiv",
   eprint = {1502.03112},
 primaryClass = "astro-ph.HE",
     year = 2015,
    month = sep,
   volume = 452,
    pages = {2540-2545},
      doi = {10.1093/mnras/stv1457},
   adsurl = {http://adsabs.harvard.edu/abs/2015MNRAS.452.2540D}
}

@ARTICLE{PG1302Nature:2015b,
   author = {{D'Orazio}, D.~J. and {Haiman}, Z. and {Schiminovich}, D.},
    title = "{Relativistic boost as the cause of periodicity in a massive black-hole binary candidate}",
  journal = {Nature},
archivePrefix = "arXiv",
   eprint = {1509.04301},
 primaryClass = "astro-ph.HE",
     year = 2015,
    month = sep,
     volume = 525,
    pages = {351-353},
   adsurl = {http://adsabs.harvard.edu/abs/2015arXiv150904301D}
}

@ARTICLE{ShiKrolik:2015,
   author = {{Shi}, J.-M. and {Krolik}, J.~H.},
    title = "{Three-dimensional MHD Simulation of Circumbinary Accretion Disks. II. Net Accretion Rate}",
  journal = {\apj},
archivePrefix = "arXiv",
   eprint = {1503.05561},
 primaryClass = "astro-ph.HE",
     year = 2015,
    month = jul,
   volume = 807,
      eid = {131},
    pages = {131},
      doi = {10.1088/0004-637X/807/2/131},
   adsurl = {http://adsabs.harvard.edu/abs/2015ApJ...807..131S}
}

@ARTICLE{Dunhill+2015,
   author = {{Dunhill}, A.~C. and {Cuadra}, J. and {Dougados}, C.},
    title = "{Precession and accretion in circumbinary discs: the case of HD 104237}",
  journal = {\mnras},
archivePrefix = "arXiv",
   eprint = {1411.0687},
 primaryClass = "astro-ph.SR",
     year = 2015,
    month = apr,
   volume = 448,
    pages = {3545-3554},
      doi = {10.1093/mnras/stv284},
   adsurl = {http://adsabs.harvard.edu/abs/2015MNRAS.448.3545D}
}

@ARTICLE{BlandfordKonigl:1979,
       author = {{Blandford}, R.~D. and {K{\"o}nigl}, A.},
        title = "{Relativistic jets as compact radio sources.}",
      journal = {\apj},
         year = 1979,
        month = aug,
       volume = {232},
        pages = {34-48},
          doi = {10.1086/157262},
       adsurl = {https://ui.adsabs.harvard.edu/abs/1979ApJ...232...34B}
}

@ARTICLE{Spada:jets:2001,
       author = {{Spada}, Maddalena and {Ghisellini}, Gabriele and {Lazzati}, Davide and {Celotti}, Annalisa},
        title = "{Internal shocks in the jets of radio-loud quasars}",
      journal = {\mnras},
         year = 2001,
        month = aug,
       volume = {325},
       number = {4},
        pages = {1559-1570},
          doi = {10.1046/j.1365-8711.2001.04557.x},
archivePrefix = {arXiv},
       eprint = {astro-ph/0103424},
 primaryClass = {astro-ph},
       adsurl = {https://ui.adsabs.harvard.edu/abs/2001MNRAS.325.1559S}
}

@ARTICLE{YuanCuiNarayan:2005,
       author = {{Yuan}, Feng and {Cui}, Wei and {Narayan}, Ramesh},
        title = "{An Accretion-Jet Model for Black Hole Binaries: Interpreting the Spectral and Timing Features of XTE J1118+480}",
      journal = {\apj},
         year = 2005,
        month = feb,
       volume = {620},
       number = {2},
        pages = {905-914},
          doi = {10.1086/427206},
archivePrefix = {arXiv},
       eprint = {astro-ph/0407612},
 primaryClass = {astro-ph},
       adsurl = {https://ui.adsabs.harvard.edu/abs/2005ApJ...620..905Y}
}

@ARTICLE{Munoz:2019,
   author = {{Mu{\~n}oz}, D.~J. and {Miranda}, R. and {Lai}, D.},
    title = "{Hydrodynamics of Circumbinary Accretion: Angular Momentum Transfer and Binary Orbital Evolution}",
  journal = {\apj},
archivePrefix = "arXiv",
   eprint = {1810.04676},
 primaryClass = "astro-ph.HE",
     year = 2019,
    month = jan,
   volume = 871,
      eid = {84},
    pages = {84},
      doi = {10.3847/1538-4357/aaf867},
   adsurl = {http://adsabs.harvard.edu/abs/2019ApJ...871...84M}
}

@article{YoungClarke:2015,
author = {Young, M D and Clarke, C J},
title = {{Binary accretion rates: dependence on temperature and mass ratio}},
journal = {Monthly Notices of the Royal Astronomical Society},
year = {2015},
volume = {452},
number = {3},
pages = {3085--3091},
month = sep
}

@article{YoungBairdClarke:2015,
author = {Young, M D and Baird, J T and Clarke, C J},
title = {{The evolution of the mass ratio of accreting binaries: the role of gas temperature}},
journal = {Monthly Notices of the Royal Astronomical Society},
year = {2015},
volume = {447},
number = {3},
pages = {2907--2914},
month = mar
}

@article{GenerozovHaiman:2014,
author = {Generozov, Aleksey and Haiman, Zoltan},
title = {{Lyman edges in supermassive black hole binaries}},
journal = {Monthly Notices of the Royal Astronomical Society: Letters},
year = {2014},
volume = {443},
number = {1},
pages = {L64--L68},
month = sep
}

@ARTICLE{Graham+2015a,
   author = {{Graham}, M.~J. and {Djorgovski}, S.~G. and {Stern}, D. and 
  {Glikman}, E. and {Drake}, A.~J. and {Mahabal}, A.~A. and {Donalek}, C. and 
  {Larson}, S. and {Christensen}, E.},
    title = "{A possible close supermassive black-hole binary in a quasar with optical periodicity}",
  journal = {\nat},
archivePrefix = "arXiv",
   eprint = {1501.01375},
     year = 2015,
    month = feb,
   volume = 518,
    pages = {74-76},
      doi = {10.1038/nature14143},
   adsurl = {http://adsabs.harvard.edu/abs/2015Natur.518...74G}
}

@ARTICLE{NarayanYi:ADAF:1994,
       author = {{Narayan}, Ramesh and {Yi}, Insu},
        title = "{Advection-dominated Accretion: A Self-similar Solution}",
      journal = {\apjl},
         year = 1994,
        month = jun,
       volume = {428},
        pages = {L13},
          doi = {10.1086/187381},
archivePrefix = {arXiv},
       eprint = {astro-ph/9403052},
 primaryClass = {astro-ph},
       adsurl = {https://ui.adsabs.harvard.edu/abs/1994ApJ...428L..13N}
}

@ARTICLE{NarayanYi:ADAF:1995,
       author = {{Narayan}, Ramesh and {Yi}, Insu},
        title = "{Advection-dominated Accretion: Underfed Black Holes and Neutron Stars}",
      journal = {\apj},
         year = 1995,
        month = oct,
       volume = {452},
        pages = {710},
          doi = {10.1086/176343},
archivePrefix = {arXiv},
       eprint = {astro-ph/9411059},
 primaryClass = {astro-ph},
       adsurl = {https://ui.adsabs.harvard.edu/abs/1995ApJ...452..710N}
}

@INPROCEEDINGS{NMQ:ADAF:1998,
   author = {{Narayan}, R. and {Mahadevan}, R. and {Quataert}, E.},
    title = "{Advection-dominated accretion around black holes}",
booktitle = {Theory of Black Hole Accretion Disks},
     year = 1998,
   editor = {{Abramowicz}, M.~A. and {Bj{\"o}rnsson}, G. and {Pringle}, J.~E.
  },
    pages = {148-182},
   adsurl = {http://adsabs.harvard.edu/abs/1998tbha.conf..148N}
}

@ARTICLE{Mahadevan:1997,
   author = {{Mahadevan}, R.},
    title = "{Scaling Laws for Advection-dominated Flows: Applications to Low-Luminosity Galactic Nuclei}",
  journal = {\apj},
     year = 1997,
    month = mar,
   volume = 477,
    pages = {585-601},
   adsurl = {http://adsabs.harvard.edu/abs/1997ApJ...477..585M}
}

@ARTICLE{Pesce:LLAGNSED:2021,
       author = {{Pesce}, Dominic W. and {Palumbo}, Daniel C.~M. and {Narayan}, Ramesh and {Blackburn}, Lindy and {Doeleman}, Sheperd S. and {Johnson}, Michael D. and {Ma}, Chung-Pei and {Nagar}, Neil M. and {Natarajan}, Priyamvada and {Ricarte}, Angelo},
        title = "{Toward Determining the Number of Observable Supermassive Black Hole Shadows}",
      journal = {\apj},
         year = 2021,
        month = dec,
       volume = {923},
       number = {2},
          eid = {260},
        pages = {260},
          doi = {10.3847/1538-4357/ac2eb5},
archivePrefix = {arXiv},
       eprint = {2108.05228},
 primaryClass = {astro-ph.HE},
       adsurl = {https://ui.adsabs.harvard.edu/abs/2021ApJ...923..260P}
}

@ARTICLE{Charisi+2016,
       author = {{Charisi}, M. and {Bartos}, I. and {Haiman}, Z. and {Price-Whelan}, A.~M. and {Graham}, M.~J. and {Bellm}, E.~C. and {Laher}, R.~R. and {M{\'a}rka}, S.},
        title = "{A population of short-period variable quasars from PTF as supermassive black hole binary candidates}",
      journal = {\mnras},
         year = 2016,
        month = dec,
       volume = {463},
       number = {2},
        pages = {2145-2171},
          doi = {10.1093/mnras/stw1838},
archivePrefix = {arXiv},
       eprint = {1604.01020},
 primaryClass = {astro-ph.GA},
       adsurl = {https://ui.adsabs.harvard.edu/abs/2016MNRAS.463.2145C}
}

@ARTICLE{LiuErac:2016,
   author = {{Liu}, J. and {Eracleous}, M. and {Halpern}, J.~P.},
    title = "{A Radial Velocity Test for Supermassive Black Hole Binaries as an Explanation for Broad, Double-peaked Emission Lines in Active Galactic Nuclei}",
  journal = {\apj},
archivePrefix = "arXiv",
   eprint = {1512.01825},
 primaryClass = "astro-ph.HE",
     year = 2016,
    month = jan,
   volume = 817,
      eid = {42},
    pages = {42},
      doi = {10.3847/0004-637X/817/1/42},
   adsurl = {http://adsabs.harvard.edu/abs/2016ApJ...817...42L}
}

@ARTICLE{EHT:M87:2019,
       author = {{Event Horizon Telescope Collaboration} and {Akiyama}, Kazunori and {Alberdi}, Antxon and {Alef}, Walter and {Asada}, Keiichi and {Azulay}, Rebecca and {Baczko}, Anne-Kathrin and {Ball}, David and {Balokovi{\'c}}, Mislav and {Barrett}, John and {Bintley}, Dan and {Blackburn}, Lindy and {Boland}, Wilfred and {Bouman}, Katherine L. and {Bower}, Geoffrey C. and {Bremer}, Michael and {Brinkerink}, Christiaan D. and {Brissenden}, Roger and {Britzen}, Silke and {Broderick}, Avery E. and {Broguiere}, Dominique and {Bronzwaer}, Thomas and {Byun}, Do-Young and {Carlstrom}, John E. and {Chael}, Andrew and {Chan}, Chi-kwan and {Chatterjee}, Shami and {Chatterjee}, Koushik and {Chen}, Ming-Tang and {Chen}, Yongjun and {Cho}, Ilje and {Christian}, Pierre and {Conway}, John E. and {Cordes}, James M. and {Crew}, Geoffrey B. and {Cui}, Yuzhu and {Davelaar}, Jordy and {De Laurentis}, Mariafelicia and {Deane}, Roger and {Dempsey}, Jessica and {Desvignes}, Gregory and {Dexter}, Jason and {Doeleman}, Sheperd S. and {Eatough}, Ralph P. and {Falcke}, Heino and {Fish}, Vincent L. and {Fomalont}, Ed and {Fraga-Encinas}, Raquel and {Freeman}, William T. and {Friberg}, Per and {Fromm}, Christian M. and {G{\'o}mez}, Jos{\'e} L. and {Galison}, Peter and {Gammie}, Charles F. and {Garc{\'\i}a}, Roberto and {Gentaz}, Olivier and {Georgiev}, Boris and {Goddi}, Ciriaco and {Gold}, Roman and {Gu}, Minfeng and {Gurwell}, Mark and {Hada}, Kazuhiro and {Hecht}, Michael H. and {Hesper}, Ronald and {Ho}, Luis C. and {Ho}, Paul and {Honma}, Mareki and {Huang}, Chih-Wei L. and {Huang}, Lei and {Hughes}, David H. and {Ikeda}, Shiro and {Inoue}, Makoto and {Issaoun}, Sara and {James}, David J. and {Jannuzi}, Buell T. and {Janssen}, Michael and {Jeter}, Britton and {Jiang}, Wu and {Johnson}, Michael D. and {Jorstad}, Svetlana and {Jung}, Taehyun and {Karami}, Mansour and {Karuppusamy}, Ramesh and {Kawashima}, Tomohisa and {Keating}, Garrett K. and {Kettenis}, Mark and {Kim}, Jae-Young and {Kim}, Junhan and {Kim}, Jongsoo and {Kino}, Motoki and {Koay}, Jun Yi and {Koch}, Patrick M. and {Koyama}, Shoko and {Kramer}, Michael and {Kramer}, Carsten and {Krichbaum}, Thomas P. and {Kuo}, Cheng-Yu and {Lauer}, Tod R. and {Lee}, Sang-Sung and {Li}, Yan-Rong and {Li}, Zhiyuan and {Lindqvist}, Michael and {Liu}, Kuo and {Liuzzo}, Elisabetta and {Lo}, Wen-Ping and {Lobanov}, Andrei P. and {Loinard}, Laurent and {Lonsdale}, Colin and {Lu}, Ru-Sen and {MacDonald}, Nicholas R. and {Mao}, Jirong and {Markoff}, Sera and {Marrone}, Daniel P. and {Marscher}, Alan P. and {Mart{\'\i}-Vidal}, Iv{\'a}n and {Matsushita}, Satoki and {Matthews}, Lynn D. and {Medeiros}, Lia and {Menten}, Karl M. and {Mizuno}, Yosuke and {Mizuno}, Izumi and {Moran}, James M. and {Moriyama}, Kotaro and {Moscibrodzka}, Monika and {M{\"u}ller}, Cornelia and {Nagai}, Hiroshi and {Nagar}, Neil M. and {Nakamura}, Masanori and {Narayan}, Ramesh and {Narayanan}, Gopal and {Natarajan}, Iniyan and {Neri}, Roberto and {Ni}, Chunchong and {Noutsos}, Aristeidis and {Okino}, Hiroki and {Olivares}, H{\'e}ctor and {Ortiz-Le{\'o}n}, Gisela N. and {Oyama}, Tomoaki and {{\"O}zel}, Feryal and {Palumbo}, Daniel C.~M. and {Patel}, Nimesh and {Pen}, Ue-Li and {Pesce}, Dominic W. and {Pi{\'e}tu}, Vincent and {Plambeck}, Richard and {PopStefanija}, Aleksandar and {Porth}, Oliver and {Prather}, Ben and {Preciado-L{\'o}pez}, Jorge A. and {Psaltis}, Dimitrios and {Pu}, Hung-Yi and {Ramakrishnan}, Venkatessh and {Rao}, Ramprasad and {Rawlings}, Mark G. and {Raymond}, Alexander W. and {Rezzolla}, Luciano and {Ripperda}, Bart and {Roelofs}, Freek and {Rogers}, Alan and {Ros}, Eduardo and {Rose}, Mel and {Roshanineshat}, Arash and {Rottmann}, Helge and {Roy}, Alan L. and {Ruszczyk}, Chet and {Ryan}, Benjamin R. and {Rygl}, Kazi L.~J. and {S{\'a}nchez}, Salvador and {S{\'a}nchez-Arguelles}, David and {Sasada}, Mahito and {Savolainen}, Tuomas and {Schloerb}, F. Peter and {Schuster}, Karl-Friedrich and {Shao}, Lijing and {Shen}, Zhiqiang and {Small}, Des and {Sohn}, Bong Won and {SooHoo}, Jason and {Tazaki}, Fumie and {Tiede}, Paul and {Tilanus}, Remo P.~J. and {Titus}, Michael and {Toma}, Kenji and {Torne}, Pablo and {Trent}, Tyler and {Trippe}, Sascha and {Tsuda}, Shuichiro and {van Bemmel}, Ilse and {van Langevelde}, Huib Jan and {van Rossum}, Daniel R. and {Wagner}, Jan and {Wardle}, John and {Weintroub}, Jonathan and {Wex}, Norbert and {Wharton}, Robert and {Wielgus}, Maciek and {Wong}, George N. and {Wu}, Qingwen and {Young}, Ken and {Young}, Andr{\'e}},
        title = "{First M87 Event Horizon Telescope Results. I. The Shadow of the Supermassive Black Hole}",
      journal = {\apjl},
         year = 2019,
        month = apr,
       volume = {875},
       number = {1},
          eid = {L1},
        pages = {L1},
          doi = {10.3847/2041-8213/ab0ec7},
archivePrefix = {arXiv},
       eprint = {1906.11238},
 primaryClass = {astro-ph.GA},
       adsurl = {https://ui.adsabs.harvard.edu/abs/2019ApJ...875L...1E}
}

@ARTICLE{KelleyHaiman:2019,
       author = {{Kelley}, Luke Zoltan and {Haiman}, Zolt{\'a}n and {Sesana}, Alberto and {Hernquist}, Lars},
        title = "{Massive BH binaries as periodically variable AGN}",
      journal = {\mnras},
         year = 2019,
        month = may,
       volume = {485},
       number = {2},
        pages = {1579-1594},
          doi = {10.1093/mnras/stz150},
archivePrefix = {arXiv},
       eprint = {1809.02138},
 primaryClass = {astro-ph.HE},
       adsurl = {https://ui.adsabs.harvard.edu/abs/2019MNRAS.485.1579K}
}

@ARTICLE{ShenLoeb:2010,
   author = {{Shen}, Y. and {Loeb}, A.},
    title = "{Identifying Supermassive Black Hole Binaries with Broad Emission Line Diagnosis}",
  journal = {\apj},
archivePrefix = "arXiv",
   eprint = {0912.0541},
 primaryClass = "astro-ph.CO",
     year = 2010,
    month = dec,
   volume = 725,
    pages = {249-260},
      doi = {10.1088/0004-637X/725/1/249},
   adsurl = {http://adsabs.harvard.edu/abs/2010ApJ...725..249S}
}

@ARTICLE{Kun+2015:PG1302,
   author = {{Kun}, E. and {Frey}, S. and {Gab{\'a}nyi}, K.~{\'E}. and {Britzen}, S. and 
	{Cseh}, D. and {Gergely}, L.~{\'A}.},
    title = "{Constraining the parameters of the putative supermassive binary black hole in PG 1302-102 from its radio structure}",
  journal = {\mnras},
archivePrefix = "arXiv",
   eprint = {1506.07036},
 primaryClass = "astro-ph.HE",
     year = 2015,
    month = dec,
   volume = 454,
    pages = {1290-1296},
      doi = {10.1093/mnras/stv2049},
   adsurl = {http://adsabs.harvard.edu/abs/2015MNRAS.454.1290K}
}

@ARTICLE{MerrittEker:2002,
   author = {{Merritt}, D. and {Ekers}, R.~D.},
    title = "{Tracing Black Hole Mergers Through Radio Lobe Morphology}",
  journal = {Science},
   eprint = {astro-ph/0208001},
     year = 2002,
    month = aug,
   volume = 297,
    pages = {1310-1313},
      doi = {10.1126/science.1074688},
   adsurl = {http://adsabs.harvard.edu/abs/2002Sci...297.1310M}
}

@ARTICLE{DDLens:2018,
       author = {{D'Orazio}, Daniel J. and {Di Stefano}, Rosanne},
        title = "{Periodic self-lensing from accreting massive black hole binaries}",
      journal = {\mnras},
         year = 2018,
        month = mar,
       volume = {474},
       number = {3},
        pages = {2975-2986},
          doi = {10.1093/mnras/stx2936},
archivePrefix = {arXiv},
       eprint = {1707.02335},
 primaryClass = {astro-ph.HE},
       adsurl = {https://ui.adsabs.harvard.edu/abs/2018MNRAS.474.2975D}
}

@ARTICLE{Phinney:2001,
   author = {{Phinney}, E.~S.},
    title = "{A Practical Theorem on Gravitational Wave Backgrounds}",
  journal = {ArXiv Astrophysics e-prints},
   eprint = {astro-ph/0108028},
     year = 2001,
    month = aug,
   adsurl = {http://adsabs.harvard.edu/abs/2001astro.ph..8028P}
}

@ARTICLE{MirandaLai+2017,
   author = {{Miranda}, R. and {Mu{\~n}oz}, D.~J. and {Lai}, D.},
    title = "{Viscous hydrodynamics simulations of circumbinary accretion discs: variability, quasi-steady state and angular momentum transfer}",
  journal = {\mnras},
archivePrefix = "arXiv",
   eprint = {1610.07263},
 primaryClass = "astro-ph.SR",
     year = 2017,
    month = apr,
   volume = 466,
    pages = {1170-1191},
      doi = {10.1093/mnras/stw3189},
   adsurl = {http://adsabs.harvard.edu/abs/2017MNRAS.466.1170M}
}

@ARTICLE{Tiede:2020,
       author = {{Tiede}, Christopher and {Zrake}, Jonathan and {MacFadyen}, Andrew and {Haiman}, Zoltan},
        title = "{Gas-driven Inspiral of Binaries in Thin Accretion Disks}",
      journal = {\apj},
         year = 2020,
        month = sep,
       volume = {900},
       number = {1},
          eid = {43},
        pages = {43},
          doi = {10.3847/1538-4357/aba432},
archivePrefix = {arXiv},
       eprint = {2005.09555},
 primaryClass = {astro-ph.GA},
       adsurl = {https://ui.adsabs.harvard.edu/abs/2020ApJ...900...43T}
}

@ARTICLE{Tiede:2022,
       author = {{Tiede}, Christopher and {Zrake}, Jonathan and {MacFadyen}, Andrew and {Haiman}, Zolt{\'a}n},
        title = "{How Binaries Accrete: Hydrodynamic Simulations with Passive Tracer Particles}",
      journal = {\apj},
         year = 2022,
        month = jun,
       volume = {932},
       number = {1},
          eid = {24},
        pages = {24},
          doi = {10.3847/1538-4357/ac6c2b},
archivePrefix = {arXiv},
       eprint = {2111.04721},
 primaryClass = {astro-ph.GA},
       adsurl = {https://ui.adsabs.harvard.edu/abs/2022ApJ...932...24T}
}

@ARTICLE{Siwek:2023,
       author = {{Siwek}, Magdalena and {Weinberger}, Rainer and {Hernquist}, Lars},
        title = "{Orbital evolution of binaries in circumbinary discs}",
      journal = {\mnras},
         year = 2023,
        month = jun,
       volume = {522},
       number = {2},
        pages = {2707-2717},
          doi = {10.1093/mnras/stad1131},
archivePrefix = {arXiv},
       eprint = {2302.01785},
 primaryClass = {astro-ph.HE},
       adsurl = {https://ui.adsabs.harvard.edu/abs/2023MNRAS.522.2707S}
}

\end{document}